\documentclass[twocolumn,prl,showpacs,superscriptaddress]{revtex4-1}
\usepackage{graphicx,color}
\usepackage{amsmath,amssymb,latexsym,amsthm}
\usepackage{mathtools}
\usepackage{mhchem}
\usepackage{tikz}
\usetikzlibrary{chains,shapes,fit,calc,arrows}
\usepackage{pgfplots, pgfplotstable}
\usepgfplotslibrary{groupplots}

\usepackage{color}

\newcommand{\mean}[1]{{\left< #1 \right>}}

\definecolor{webgreen}{rgb}{0,.5,0}
\definecolor{webbrown}{rgb}{.6,0,0}
\definecolor{grigio}{rgb}{.85,.85,.85} 
\definecolor{RoyalBlue}{rgb}{0.0, 0.14, 0.4}
\definecolor{skyblue1}{rgb}{0.45,0.62,0.81}
\definecolor{skyblue2}{rgb}{0.2,0.39,0.64}
\definecolor{skyblue3}{rgb}{0.13,0.29,0.53}
\definecolor{scarlet1}{rgb}{0.93,0.16,0.16}
\definecolor{scarlet2}{rgb}{0.8,0,0}
\definecolor{scarlet3}{rgb}{0.64,0,0}

\usepackage{hyperref}
\hypersetup{%
    colorlinks=true, linktocpage=true, pdfstartpage=1, pdfstartview=FitV,%
    breaklinks=true, pdfpagemode=UseNone, pageanchor=true, pdfpagemode=UseOutlines,%
    plainpages=false, bookmarksnumbered, bookmarksopen=true, bookmarksopenlevel=1,%
    hypertexnames=true, pdfhighlight=/O,
    urlcolor=webbrown, linkcolor=RoyalBlue, citecolor=webgreen, 
    pdftitle={Information Thermodynamics of Turing Patterns},%
    pdfauthor={Gianmaria Falasco},%
    pdfsubject={},%
    pdfkeywords={},%
    pdfcreator={pdfLaTeX},%
    pdfproducer={LaTeX REVTeX}%
}
\begin{document}

\title{Unifying Thermodynamic Uncertainty Relations}
\newcommand\unilux{\affiliation{Complex Systems and Statistical Mechanics, Physics and Materials Science Research Unit, University of Luxembourg, L-1511 Luxembourg}}
\newcommand\unilou{\affiliation{Institute of Information and Communication Technologies, Electronics and Applied Mathematics Universit\'{e} catholique de Louvain, Louvain-La-Neuve, Belgium
}}
\author{Gianmaria Falasco}
\unilux
\author{Massimiliano Esposito}
\unilux
\author{Jean-Charles Delvenne}
\unilou

\begin{abstract}

We introduce a new technique to bound the fluctuations exhibited by a physical system, based on the Euclidean geometry of the space of observables. Through a simple unifying argument, we derive a sweeping generalization of so-called Thermodynamic Uncertainty Relations (TURs). We not only strengthen the bounds but extend their realm of applicability and in many cases prove their optimality, without resorting to Large Deviation theory or information-theoretic techniques. In particular, we find the best TUR based on entropy production alone and also derive a novel bound for stationary Markov processes, which surpasses previous known bounds. Our results derive from the non-invariance of the system under a symmetry which can be other than time reversal and thus open a wide new spectrum of applications.  
\end{abstract}


\maketitle

At several levels of complexity, random processes are successfully employed to model  natural phenomena, such as  open quantum system \cite{bre02}, soft and active matter \cite{fre05}, biochemical reactions \cite{bre14}, and population ecology \cite{ova09}, just to name a few. In recent years, the understanding of their dynamical fluctuations has greatly advanced thanks to exact results of nonequilibrium physics. 
Most importantly, fluctuation theorems \cite{esp09, rao18} and response relations \cite{bai13} have been derived that, respectively, constrain the distribution of currents and relate the system's perturbation to its dissipation and dynamical activity. Moreover, stochastic thermodynamics has emerged as a comprehensive framework to rigorously study the energetics and thermodynamics of stochastic processes \cite{sei12, rao18b}.

Recently, uncertainty relations appeared as a new powerful tool to investigate dynamical fluctuations.
They denote a set of inequalities in which the square-mean-to-variance ratio, or precision $\mathfrak{p}(f)$, of a generic observable $f$ integrated over a time interval $t_{\text{f}}$ is bounded by an $f$-independent functional $\mathfrak{p}_\text{max}$:
\begin{align}\label{prec}
\mathfrak{p}(f):=\frac{|\langle f \rangle|^2}{\text{Var}f} \leqslant  \mathfrak{p}_\text{max} \,.
\end{align}
It was first conjectured in \cite{bar15}  that $\mathfrak{p}$ for a time-integrated current-like (i.e. odd under time reversal) observable $f$ is bounded by half the expected entropy $\langle \sigma \rangle$ produced over the interval $t_{\text{f}}$, i.e. $\mathfrak{p}_\text{max}\leq\langle \sigma \rangle/2$. This so-called thermodynamic uncertainty relation, originally proved in the linear response regime and under stationary conditions, triggered an intense activity seeking generalizations or improvements for the largest possible class of out-of-equilibrium conditions. Apart from its conceptual importance, i.e. the existence of an universal upper bound set by dissipation on the precision of any current, \eqref{prec} has major practical consequences. Indeed, \eqref{prec} allows one to bound functions of the system's dissipation which are not directly measurable, e.g. the thermodynamic efficiency of molecular motors \cite{pie16}, or to reveal the existence of hidden nonequilibrium states \cite{how19}. A first proof valid beyond the linear regime but restricted to large time intervals $t_{\text{f}}$ \cite{gin16} was soon extended to arbitrary $t_{\text{f}}$ \cite{hor17}. These, and related early results \cite{pie16a,pol16,gin17,nar17,mae17} were obtained within large deviation theory, by progressively refining the bound on the rate function for empirical currents of jump and diffusion processes. Simultaneously, the same formalism was employed to extend \eqref{prec} to counting observables of jump processes \cite{gar17}. In this context it was found that $\mathfrak{p}$ is bounded by the mean of the total number of jumps, or  activity, occurring in the time span $t_{\text{f}}$.

\vspace{-2pt}
A different method to tackle the problem, based on perturbing the generating function of an arbitrary observable $f$, was designed in \cite{DS18}. It yields an upper bound for the response of $f$, which reduces to \eqref{prec} when the chosen perturbation results in a time rescaling of the dynamics. The entropic \cite{dec18} as well as the activity bound \cite{dit18} have thus been extended to both current-like and counting observables. 
This approach, which makes contact with inequalities originally derived by Kullback \cite{kul}, has sparked much interest in the application of information theoretic results and concepts. 

\vspace{-2pt}
More recently \cite{has19}, the exponential bound $\mathfrak{p}_\text{max}=(\exp \langle \sigma \rangle -1)/2$ has been derived for Langevin dynamics with feedback, under the condition of validity of the detailed (joint) fluctuation theorem for $\sigma$ and $f$.
The same bound had already been derived in~\cite{pro17} for periodically driven Markovian systems with a time-symmetric protocol, where now $\langle \sigma \rangle$ is computed over one period and $(\exp \langle \sigma \rangle -1)/2$ bounds the precision divided by the (asymptotically large) number of  periods.

\vspace{-2pt}
Here, we provide an overarching method, based on elementary observations on the Hilbert space structure of observables, to recover and generalize the various bounds obtained so far in the literature. First, we provide an exact expression for $\mathfrak{p}_\text{max}$ in the case of arbitrary stochastic processes, possibly non-Markovian, time-varying or non-stationary, and show that the bound $(\exp \langle \sigma \rangle -1)/2$ can be improved by a factor 2, and no more. In the case of periodic Markovian processes, we show that the precision over a period bounds the precision per period over arbitrary time intervals, which trivializes all the asymptotic bounds obtained so far in the periodic Markovian case. In the case of stationary time-invariant Markov processes, it also allows to replace them with simple and tighter bounds, valid over all time intervals.

\emph{The Hilbert Uncertainty Relation---} We first state the most abstract version of our result. We consider a general real or complex Hilbert space $\mathcal{F}$ with  some scalar product $\langle . | . \rangle$. To every $f \in \mathcal{F}$ is associated the so-called mean value of $f$, a scalar quantity $\mean{f}$ that is linear and continuous in $f$, i.e. a one-form in the dual of $\mathcal{F}$. By virtue of the Riesz representation theorem, one can find a special element $m$ in $\mathcal{F}$, so that the mean is expressed as $\mean{f}= \langle m | f \rangle, \forall f \in \mathcal{F}$.
We call $m$ an averaging observable for $\mathcal{F}$.  
We now consider the following ratio, that we called normalized precision for reasons that will appear clearly below,
\begin{align}\label{np}
\mathfrak{np}(f) := \frac{|\langle f \rangle| ^2}{\langle f | f \rangle}. 
\end{align}
Through Cauchy-Schwarz inequality we get
$| \mean{f} |^2 = |\langle m | f \rangle|^2 \leq \langle m |m \rangle \langle f | f \rangle$,
with equality when $f$ is aligned with $m$. Thus
\begin{align}\label{npmax}
\mathfrak{np}_\text{max}:=\max_{f \in \mathcal{F}} \frac{|\mean{f}|^2}{ \langle f | f \rangle} = \langle m |m \rangle = \mean{m}.
\end{align}
This constitutes the key observation of this article which we call the Hilbert Uncertainty Relation.

To be concrete, we focus on classical physical systems described by a configuration space $\Omega$ whose elements $\omega$ are, for example, trajectories of a random dynamical system. The configuration space is endowed with a probability measure $p(\omega)$.  An obvious Hilbert space of interest is  the space $\mathcal{L}^2(\Omega)$ of square-summable observables, i.e. functions $f: \Omega \to \mathbb{R}$ such that the mean $\langle f \rangle = \sum_\omega f(\omega) p(\omega) $ and the mean square $\langle f |f \rangle =  \sum_\omega f(\omega)^2 p(\omega) $ are well-defined and finite (even though our considerations also apply to continuous cases, we adopt the discrete summation notations). The normalized precision now ranges between zero and one, and is equivalent to precision via the relation $\mathfrak{p}(f)=\mathfrak{np}(f)/(1-\mathfrak{np}(f))$.  
In this situation, the averaging observable $m$ is simply the constant observable 1, so that $\mathfrak{np}_\text{max}=1$, corresponding to zero variance and infinite precision. 

However in many situations we are interested in a (closed) linear subspace $\mathcal{F}$ of those observables, sharing some properties of interest, which we call for the sake of convenience the `legitimate observables'. If this subspace, itself a Hilbert space for the same scalar product, does not contain the constant observables, then there is  a non-trivial legitimate averaging observable $m$, for which $\mathfrak{np}_\text{max}$ now caps the normalized precision of all legitimate observables. It is also the orthogonal projection of the constant observable 1 onto the space $\mathcal{F}$ of legitimate observables, as $\mean{f}= \langle 1|f \rangle = \langle m|f \rangle$ implies that $1-m$ is orthogonal to all legitimate observables. The corresponding $\mathfrak{p}_\text{max}$ over $\mathcal{F}$ is $\langle m | m \rangle/(1-\langle m | m \rangle)$.
 
\begin{figure} 
	\centering
	\includegraphics[width=0.5\textwidth]{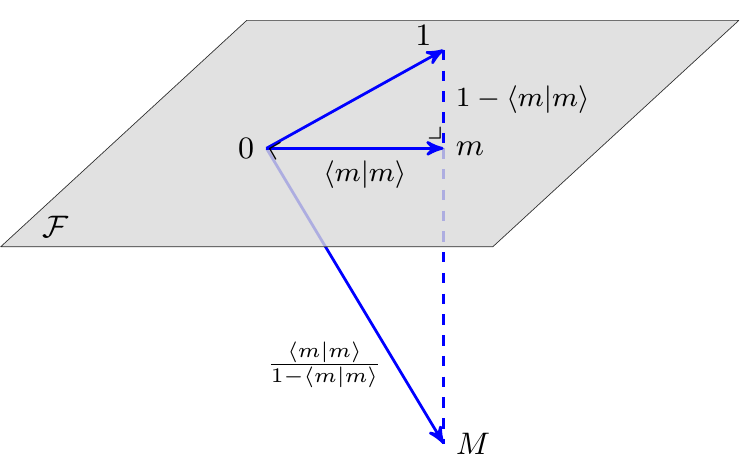}
	\caption[]{(Left) Three averaging observables for the legitimate observables $\mathcal{F}$: the unit observable, the legitimate observable $m$ and the zero-mean observable $M$. The square lengths reported on the diagram show via elementary arguments on similar Pythagorean triangles that $\langle M | M \rangle$ is indeed the maximum precision $\langle m | m \rangle/(1-\langle m | m \rangle)$.}
	\label{fig:proof}
\end{figure}

Interestingly this quantity has a geometric interpretation. Assume that we find a zero-mean square-summable observable $M$---possibly illegitimate, i.e. outside of $\mathcal{F}$---that is still an averaging observable, i.e. verifying $\langle M|f \rangle=\langle f \rangle$ for all legitimate observables $f$. Then $\langle M|f \rangle$ is also the covariance of $M$ with $f$, since $\langle M \rangle =0$, and $\langle M|M \rangle$ is also the variance of $M$. Therefore, Cauchy-Schwarz inequality applied to the covariance,
 $ |\langle f \rangle|^2 = |\text{Cov}(M,f)|^2 \leq \text{Var}(M) \text{Var}(f) =  \langle M|M \rangle \text{Var}(f)$,
yields that $\langle M|M \rangle$  is an upper bound on the maximum precision of legitimate observables. In fact, if $M$ is aligned with 1 and $m$ while being orthogonal to 1, namely,
\begin{equation}\label{M}
M(\omega)=1 - \frac{1-m(\omega)}{1-\langle m \rangle},
\end{equation}
 we find that $\langle M  | M \rangle$ is exactly the maximum precision $\langle m | m \rangle/(1-\langle m | m \rangle)$ reachable over $\mathcal{F}$ (see Fig.~\eqref{fig:proof} for a geometric representation).

\emph{Time anti-symmetric observables---} The TURs are obtained by considering $\Omega$ as the set of all possible paths of a random process, endowed with an involution symmetry (i.e., a transformation whose square is the identity) called time-reversal, which maps any path $\omega$ in $\Omega$ to its time-reversed path $\overline{\omega}$. We consider legitimate the observables that are time-antisymmetric, i.e. satisfying $f(\omega)=-f(\overline{\omega})$. The time-reversal induces another probability measure $\overline{p}(\omega) = p(\overline{\omega}) $, attributing to an event the $p$-probability of the time-reversed event.  Then the scalar product of two time-antisymmetric observables $f$ and $g$ can be written as  $ \langle f | g \rangle =   \sum_\omega  f(\omega) g(\omega) (p(\omega) + \overline{p}(\omega))/2$ while the mean of $f$ is written as
$\langle f  \rangle = \sum_\omega  f(\omega)  ( p(\omega) - \overline{p}(\omega))/2$. From this we deduce that the mean observable $m$ satisfying $\langle m | f\rangle = \langle f \rangle$ is the time-antisymmetric observable:
\begin{equation}\label{general-m}
m= \frac{p - \overline{p}}{p +  \overline{p}}.
\end{equation}
The maximum normalised precision \eqref{npmax} over all time-antisymmetric observables is therefore 
\begin{align}\label{general-np}
\mathfrak{np}_\text{max}=\mean{ \frac{p -  \overline{p}}{ p +  \overline{p} }} = \frac{1}{2} \sum_\omega \frac{ (p -  \overline{p}  )^2  }{  p+ \overline{p}   } .
\end{align}
This \emph{exact} bound can be written in terms of $\sigma:=\ln \frac{p}{\overline{p}}$ as
\begin{align}\label{tanh}
\mathfrak{np}_\text{max}= \langle \tanh \frac{\sigma}{2} \rangle.
\end{align}
Equation \eqref{general-np} clearly cancels when the probability measure is time-symmetric, $p=\overline{p}$, and can be loosened in terms of two different quantities that capture the gap separating $p$ from $\overline{p}$. First, the total variation distance, ranging between zero and one, $d:= \frac{1}{2} \sum_\omega |p- \overline{p}| $. Second, the Kullback-Leibler (KL) divergence $\mean{\sigma}$. Rewriting \eqref{general-np} as $\mathfrak{np}_\text{max}/d=  \sum_\omega (|p - \overline{p}|/2d) \tanh (|\sigma|/2) $, a convex combination of positive values of the concave function $\tanh$, we obtain the relaxed inequality
\begin{equation}\label{eq:general}
\mathfrak{np}_\text{max} \leq \ d \tanh \frac{\langle\sigma\rangle}{2d},
\end{equation}
the main result of this section.
As this expression is increasing in $d$, one can use the coarse bound $d \leq 1$ to obtain $
\mathfrak{np}_\text{max} \leq \ \tanh \frac{\langle\sigma\rangle}{2}$, which in term of square-mean-to-variance ratio reads 
\begin{equation}
\label{eq:vdb}
\mathfrak{p}_\text{max} \leq \ \frac{e^{\langle \sigma  \rangle}-1}{2},
\end{equation} 
a bound recently proposed under the name of General TUR \cite{has19,pot19}. We underline that this result is valid for arbitrary dynamics, such as non-Markovian and non-autonomous,  for all time intervals $T$, and all possible time-antisymmetric observables---not necessarily time-integrated ones. It is even valid for set of paths of variable length, e.g., defined by a random stopping time. It is also valid for any notion of `time-reversal' that is an involution of $\Omega$. For example, if a path is defined as a discrete or continuous list of `states', then the time-reversed path may be defined as the time-reversed list of the same states, or the time-reversed list of conjugated states. Typically, in a model of an underdamped system we want to include the speed or momentum as part of the state, and flip it as well as reversing the order of states when applying time-reversal. All these choices for the time-reversal involution will yield mathematically valid inequalities, but not all will carry the same physical meaning. For instance, it is only in the circumstances where the fluctuation relation holds \cite{rao18} that $\mean{\sigma}$ is the physical entropy production associated with the process (as requested in \cite{has19}). One such circumstance is when the system is driven by a time-symmetric protocol, and respects local detailed balance at all times. Outside these examples, $\mean{\sigma}$ is to be regarded as an observable of interest, accessible in principle to the measurement, bearing no direct connection to thermodynamics, yet useful as a bound on the fluctations of time-antisymmetric observables such as total displacement, etc.
Note that some observables of practical interest, such as work or heat, are dependent on the parameters of the protocol and therefore are time-antisymmetric if the  time-varying protocol is itself time-symmetric. Time-symmetric protocol is an assumption requested by \cite{pro17,pot19}. In the case of arbitrary time-varying protocols, another notion of time-reversal is needed, which also reverses the protocol, in order to include those observables of interest in the space of legitimate observables. This was first investigated in \cite{pro19} with a tailored large deviation argument (see SI for a formal statement and a proof as a direct corollary of \eqref{eq:vdb}).

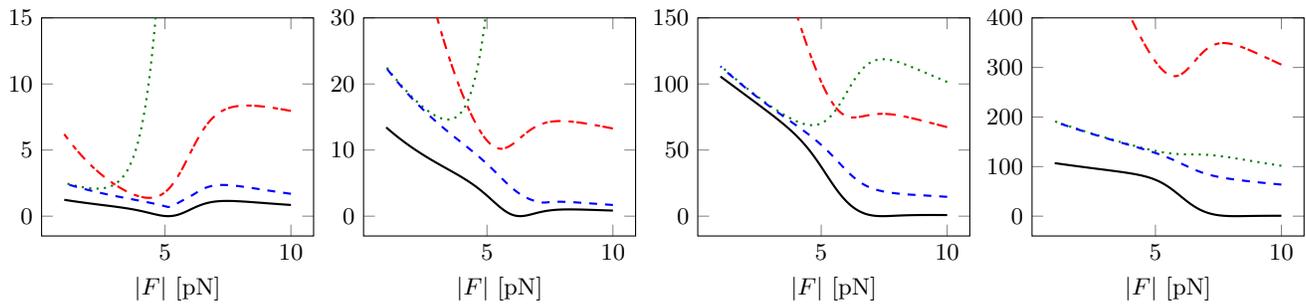
\begin{figure*}[t]
	\centering
	\begin{tikzpicture}
	\begin{axis}[width=0.29\textwidth, ymax=15, xlabel={$|F|$ [pN]},
	xlabel near ticks, ylabel near ticks, ylabel shift = -4 pt] 
	\addplot[color=black, no markers,style={thick}] coordinates{(10,0.846109 )(99/10,0.859157 )(49/5,0.872379 )(97/10,0.885771 )(48/5,0.89933 )(19/2,0.913046 )(47/5,0.926912 )(93/10,0.940915 )(46/5,0.955039 )(91/10,0.969265 )(9,0.983568 )(89/10,0.997917 )(44/5,1.01227 )(87/10,1.02659 )(43/5,1.0408 )(17/2,1.05485 )(42/5,1.06863 )(83/10,1.08205 )(41/5,1.09496 )(81/10,1.10721 )(8,1.11861 )(79/10,1.12892 )(39/5,1.13786 )(77/10,1.14511 )(38/5,1.15028 )(15/2,1.1529 )(37/5,1.15247 )(73/10,1.14837 )(36/5,1.13993 )(71/10,1.12643 )(7,1.10708 )(69/10,1.08109 )(34/5,1.04768 )(67/10,1.00617 )(33/5,0.956056 )(13/2,0.897094 )(32/5,0.829417 )(63/10,0.753631 )(31/5,0.670902 )(61/10,0.582991 )(6,0.492232 )(59/10,0.401436 )(29/5,0.313714 )(57/10,0.232248 )(28/5,0.160022 )(11/2,0.0995717 )(27/5,0.0527626 )(53/10,0.0206566 )(26/5,0.00346056 )(51/10,0.000570007 )(5,0.0106923 )(49/10,0.0320267 )(24/5,0.062471 )(47/10,0.0998265 )(23/5,0.141975 )(9/2,0.18701 )(22/5,0.233322 )(43/10,0.279635 )(21/5,0.324998 )(41/10,0.368763 )(4,0.410534 )(39/10,0.450118 )(19/5,0.487478 )(37/10,0.522685 )(18/5,0.555884 )(7/2,0.587263 )(17/5,0.617034 )(33/10,0.645413 )(16/5,0.672615 )(31/10,0.698843 )(3,0.724285 )(29/10,0.749112 )(14/5,0.773481 )(27/10,0.79753 )(13/5,0.821382 )(5/2,0.845146 )(12/5,0.868918 )(23/10,0.892784 )(11/5,0.916817 )(21/10,0.941085 )(2,0.965646 )(19/10,0.990553 )(9/5,1.01585 )(17/10,1.04159 )(8/5,1.06779 )(3/2,1.09451 )(7/5,1.12177 )(13/10,1.14959 )(6/5,1.17802 )(11/10,1.20707 )(1,1.23678)
	};
	\addplot[color=webgreen, no markers,style={thick},dotted] coordinates{
		(10,101.535 )(99/10,102.326 )(49/5,103.12 )(97/10,103.919 )(48/5,104.72 )(19/2,105.525 )(47/5,106.331 )(93/10,107.138 )(46/5,107.945 )(91/10,108.749 )(9,109.551 )(89/10,110.347 )(44/5,111.134 )(87/10,111.911 )(43/5,112.672 )(17/2,113.414 )(42/5,114.13 )(83/10,114.815 )(41/5,115.461 )(81/10,116.058 )(8,116.595 )(79/10,117.06 )(39/5,117.435 )(77/10,117.704 )(38/5,117.844 )(15/2,117.832 )(37/5,117.64 )(73/10,117.236 )(36/5,116.585 )(71/10,115.649 )(7,114.39 )(69/10,112.765 )(34/5,110.735 )(67/10,108.265 )(33/5,105.326 )(13/2,101.898 )(32/5,97.9767 )(63/10,93.5756 )(31/5,88.7276 )(61/10,83.4871 )(6,77.9285 )(59/10,72.1436 )(29/5,66.2358 )(57/10,60.3139 )(28/5,54.4845 )(11/2,48.8452 )(27/5,43.4792 )(53/10,38.4515 )(26/5,33.8077 )(51/10,29.5738 )(5,25.759 )(49/10,22.3578 )(24/5,19.3535 )(47/10,16.7218 )(23/5,14.433 )(9/2,12.4552 )(22/5,10.7556 )(43/10,9.30231 )(21/5,8.06514 )(41/10,7.01616 )(4,6.13003 )(39/10,5.38411 )(19/5,4.75839 )(37/10,4.23537 )(18/5,3.79985 )(7/2,3.43869 )(17/5,3.14065 )(33/10,2.89609 )(16/5,2.69681 )(31/10,2.53585 )(3,2.40732 )(29/10,2.30622 )(14/5,2.22835 )(27/10,2.17017 )(13/5,2.1287 )(5/2,2.10143 )(12/5,2.08625 )(23/10,2.0814 )(11/5,2.08542 )(21/10,2.09706 )(2,2.11528 )(19/10,2.13924 )(9/5,2.16821 )(17/10,2.20159 )(8/5,2.23889 )(3/2,2.2797 )(7/5,2.32368 )(13/10,2.37056 )(6/5,2.4201 )(11/10,2.47212 )(1,2.52648)
	};
	\addplot[color=red, no markers,style={thick},dash pattern={on 4pt off 1pt on 2pt off 3pt}] coordinates{
		(10,7.95849 )(99/10,7.99468 )(49/5,8.02994 )(97/10,8.06421 )(48/5,8.0974 )(19/2,8.12944 )(47/5,8.16022 )(93/10,8.18964 )(46/5,8.21756 )(91/10,8.24384 )(9,8.26831 )(89/10,8.29077 )(44/5,8.31099 )(87/10,8.3287 )(43/5,8.3436 )(17/2,8.35532 )(42/5,8.36344 )(83/10,8.36746 )(41/5,8.3668 )(81/10,8.36078 )(8,8.34863 )(79/10,8.32943 )(39/5,8.30215 )(77/10,8.26557 )(38/5,8.21834 )(15/2,8.15893 )(37/5,8.08563 )(73/10,7.99659 )(36/5,7.88982 )(71/10,7.76326 )(7,7.6148 )(69/10,7.44249 )(34/5,7.24457 )(67/10,7.0197 )(33/5,6.76717 )(13/2,6.48708 )(32/5,6.18056 )(63/10,5.84994 )(31/5,5.49882 )(61/10,5.13206 )(6,4.75559 )(59/10,4.37615 )(29/5,4.00085 )(57/10,3.63676 )(28/5,3.29039 )(11/2,2.96733 )(27/5,2.67198 )(53/10,2.40737 )(26/5,2.17521 )(51/10,1.97597 )(5,1.80909 )(49/10,1.67322 )(24/5,1.56643 )(47/10,1.48648 )(23/5,1.43095 )(9/2,1.39741 )(22/5,1.38353 )(43/10,1.38714 )(21/5,1.40627 )(41/10,1.43918 )(4,1.48433 )(39/10,1.54043 )(19/5,1.60636 )(37/10,1.68121 )(18/5,1.7642 )(7/2,1.85473 )(17/5,1.9523 )(33/10,2.05652 )(16/5,2.1671 )(31/10,2.28382 )(3,2.40654 )(29/10,2.53517 )(14/5,2.66966 )(27/10,2.81001 )(13/5,2.95625 )(5/2,3.10845 )(12/5,3.2667 )(23/10,3.4311 )(11/5,3.60178 )(21/10,3.7789 )(2,3.96261 )(19/10,4.15309 )(9/5,4.35052 )(17/10,4.55511 )(8/5,4.76706 )(3/2,4.9866 )(7/5,5.21394 )(13/10,5.44933 )(6/5,5.69301 )(11/10,5.94524 )(1,6.20629)
	};
	       \addplot[color=blue, no markers,style={thick}, dashed]  coordinates{
	(10.,1.69244)(9.9,1.71859)(9.8,1.74511)(9.7,1.77198)(9.6,1.79921)(9.5,1.82677)(9.4,1.85467)(9.3,1.88288)(9.2,1.91137)(9.1,1.94013)(9.,1.96912)(8.9,1.99828)(8.8,2.02756)(8.7,2.0569)(8.6,2.08619)(8.5,2.11534)(8.4,2.14422)(8.3,2.17265)(8.2,2.20044)(8.1,2.22735)(8.,2.2531)(7.9,2.27732)(7.8,2.29961)(7.7,2.31947)(7.6,2.33632)(7.5,2.34946)(7.4,2.35812)(7.3,2.36137)(7.2,2.3582)(7.1,2.34747)(7.,2.32796)(6.9,2.29837)(6.8,2.2574)(6.7,2.2038)(6.6,2.13647)(6.5,2.05451)(6.4,1.95743)(6.3,1.84518)(6.2,1.71826)(6.1,1.60161)(6.,1.50631)(5.9,1.40505)(5.8,1.29977)(5.7,1.19252)(5.6,1.08542)(5.5,0.980443)(5.4,0.879339)(5.3,0.78354)(5.2,0.69412)(5.1,0.700613)(5.,0.75104)(4.9,0.799524)(4.8,0.846078)(4.7,0.890783)(4.6,0.933777)(4.5,0.97523)(4.4,1.01534)(4.3,1.05429)(4.2,1.0923)(4.1,1.12956)(4.,1.16624)(3.9,1.20253)(3.8,1.23857)(3.7,1.27451)(3.6,1.31047)(3.5,1.34659)(3.4,1.38295)(3.3,1.41965)(3.2,1.45678)(3.1,1.49442)(3.,1.53263)(2.9,1.57147)(2.8,1.61101)(2.7,1.65129)(2.6,1.69236)(2.5,1.73427)(2.4,1.77707)(2.3,1.82078)(2.2,1.86546)(2.1,1.91113)(2.,1.95784)(1.9,2.00561)(1.8,2.05449)(1.7,2.1045)(1.6,2.15568)(1.5,2.20806)(1.4,2.26168)(1.3,2.31657)(1.2,2.37276)(1.1,2.43029)(1.,2.48918)
	};
	\end{axis}
	\end{tikzpicture}
	\begin{tikzpicture}
	\begin{axis}[width=0.29\textwidth, ymax=30, xlabel={$|F|$ [pN]},
	xlabel near ticks, ylabel near ticks, ylabel shift = -4 pt] 
	\addplot[color=black, no markers,style={thick}]  coordinates{(10,0.844262 )(99/10,0.856882 )(49/5,0.869577 )(97/10,0.882322 )(48/5,0.895082 )(19/2,0.907817 )(47/5,0.920476 )(93/10,0.932994 )(46/5,0.945292 )(91/10,0.957276 )(9,0.968825 )(89/10,0.979795 )(44/5,0.990009 )(87/10,0.999252 )(43/5,1.00726 )(17/2,1.01373 )(42/5,1.01828 )(83/10,1.02046 )(41/5,1.01975 )(81/10,1.01553 )(8,1.00709 )(79/10,0.993643 )(39/5,0.974288 )(77/10,0.948075 )(38/5,0.914021 )(15/2,0.871179 )(37/5,0.818725 )(73/10,0.756095 )(36/5,0.683153 )(71/10,0.600402 )(7,0.509218 )(69/10,0.412078 )(34/5,0.312752 )(67/10,0.216376 )(33/5,0.129372 )(13/2,0.059152 )(32/5,0.0135962 )(63/10,0.00035524 )(31/5,0.0260687 )(61/10,0.0956397 )(6,0.211699 )(59/10,0.374358 )(29/5,0.581268 )(57/10,0.827965 )(28/5,1.10839 )(11/2,1.41551 )(27/5,1.74194 )(53/10,2.08045 )(26/5,2.42445 )(51/10,2.76823 )(5,3.1072 )(49/10,3.43784 )(24/5,3.75777 )(47/10,4.06555 )(23/5,4.36058 )(9/2,4.64293 )(22/5,4.91317 )(43/10,5.17219 )(21/5,5.42114 )(41/10,5.66129 )(4,5.89395 )(39/10,6.12042 )(19/5,6.34196 )(37/10,6.55974 )(18/5,6.77486 )(7/2,6.98834 )(17/5,7.20108 )(33/10,7.4139 )(16/5,7.62755 )(31/10,7.8427 )(3,8.05994 )(29/10,8.27982 )(14/5,8.50282 )(27/10,8.72938 )(13/5,8.9599 )(5/2,9.19476 )(12/5,9.43428 )(23/10,9.67878 )(11/5,9.92856 )(21/10,10.1839 )(2,10.445 )(19/10,10.7122 )(9/5,10.9858 )(17/10,11.2658 )(8/5,11.5527 )(3/2,11.8465 )(7/5,12.1476 )(13/10,12.4561 )(6/5,12.7723 )(11/10,13.0964 )(1,13.4286)
	};
	\addplot[color=webgreen, no markers,style={thick},dotted] coordinates{(10,101.536 )(99/10,102.326 )(49/5,103.121 )(97/10,103.92 )(48/5,104.722 )(19/2,105.527 )(47/5,106.333 )(93/10,107.141 )(46/5,107.948 )(91/10,108.754 )(9,109.556 )(89/10,110.353 )(44/5,111.142 )(87/10,111.92 )(43/5,112.683 )(17/2,113.427 )(42/5,114.146 )(83/10,114.835 )(41/5,115.485 )(81/10,116.087 )(8,116.63 )(79/10,117.102 )(39/5,117.487 )(77/10,117.767 )(38/5,117.921 )(15/2,117.925 )(37/5,117.753 )(73/10,117.374 )(36/5,116.754 )(71/10,115.856 )(7,114.642 )(69/10,113.074 )(34/5,111.114 )(67/10,108.73 )(33/5,105.894 )(13/2,102.593 )(32/5,98.8244 )(63/10,94.6056 )(31/5,89.9728 )(61/10,84.983 )(6,79.7119 )(59/10,74.2512 )(29/5,68.7026 )(57/10,63.1712 )(28/5,57.7586 )(11/2,52.5565 )(27/5,47.6412 )(53/10,43.0711 )(26/5,38.8853 )(51/10,35.1047 )(5,31.7341 )(49/10,28.7648 )(24/5,26.1781 )(47/10,23.9488 )(23/5,22.047 )(9/2,20.4413 )(22/5,19.1 )(43/10,17.9924 )(21/5,17.0899 )(41/10,16.3661 )(4,15.7972 )(39/10,15.3621 )(19/5,15.0424 )(37/10,14.8218 )(18/5,14.6863 )(7/2,14.6241 )(17/5,14.6248 )(33/10,14.6797 )(16/5,14.7814 )(31/10,14.9238 )(3,15.1016 )(29/10,15.3103 )(14/5,15.5463 )(27/10,15.8066 )(13/5,16.0885 )(5/2,16.39 )(12/5,16.7092 )(23/10,17.0449 )(11/5,17.3957 )(21/10,17.7608 )(2,18.1393 )(19/10,18.5306 )(9/5,18.9344 )(17/10,19.3501 )(8/5,19.7775 )(3/2,20.2165 )(7/5,20.6668 )(13/10,21.1285 )(6/5,21.6014 )(11/10,22.0856 )(1,22.5811)
	};
	\addplot[color=red, no markers,style={thick},dash pattern={on 4pt off 1pt on 2pt off 3pt}] coordinates{(10,13.2572 )(99/10,13.3333 )(49/5,13.4086 )(97/10,13.4831 )(48/5,13.5565 )(19/2,13.6289 )(47/5,13.6999 )(93/10,13.7695 )(46/5,13.8375 )(91/10,13.9036 )(9,13.9676 )(89/10,14.0291 )(44/5,14.0878 )(87/10,14.1433 )(43/5,14.195 )(17/2,14.2424 )(42/5,14.2848 )(83/10,14.3214 )(41/5,14.3514 )(81/10,14.3737 )(8,14.3872 )(79/10,14.3904 )(39/5,14.382 )(77/10,14.3602 )(38/5,14.3232 )(15/2,14.2691 )(37/5,14.1956 )(73/10,14.1008 )(36/5,13.9825 )(71/10,13.8387 )(7,13.6679 )(69/10,13.4691 )(34/5,13.242 )(67/10,12.9874 )(33/5,12.7077 )(13/2,12.4067 )(32/5,12.0903 )(63/10,11.7662 )(31/5,11.4439 )(61/10,11.1345 )(6,10.8502 )(59/10,10.6033 )(29/5,10.4057 )(57/10,10.2682 )(28/5,10.1995 )(11/2,10.2062 )(27/5,10.2924 )(53/10,10.4596 )(26/5,10.7074 )(51/10,11.0337 )(5,11.435 )(49/10,11.907 )(24/5,12.4454 )(47/10,13.0455 )(23/5,13.703 )(9/2,14.414 )(22/5,15.1751 )(43/10,15.9834 )(21/5,16.8367 )(41/10,17.733 )(4,18.6711 )(39/10,19.65 )(19/5,20.6693 )(37/10,21.7286 )(18/5,22.8282 )(7/2,23.9682 )(17/5,25.1493 )(33/10,26.372 )(16/5,27.6373 )(31/10,28.9461 )(3,30.2994 )(29/10,31.6983 )(14/5,33.1442 )(27/10,34.6383 )(13/5,36.1819 )(5/2,37.7764 )(12/5,39.4234 )(23/10,41.1243 )(11/5,42.8806 )(21/10,44.6939 )(2,46.566 )(19/10,48.4984 )(9/5,50.4928 )(17/10,52.551 )(8/5,54.6749 )(3/2,56.8661 )(7/5,59.1267 )(13/10,61.4584 )(6/5,63.8633 )(11/10,66.3434 )(1,68.9007)};
   \addplot[color=blue, no markers,style={thick}, dashed]  coordinates{	
	(10.,1.6912)(9.9,1.71707)(9.8,1.74323)(9.7,1.76967)(9.6,1.79635)(9.5,1.82326)(9.4,1.85034)(9.3,1.87754)(9.2,1.9048)(9.1,1.93204)(9.,1.95915)(8.9,1.98601)(8.8,2.01245)(8.7,2.0383)(8.6,2.0633)(8.5,2.08717)(8.4,2.10956)(8.3,2.13002)(8.2,2.14804)(8.1,2.16297)(8.,2.17403)(7.9,2.18029)(7.8,2.18063)(7.7,2.17371)(7.6,2.15796)(7.5,2.13151)(7.4,2.09221)(7.3,2.05642)(7.2,2.06108)(7.1,2.09858)(7.,2.18928)(6.9,2.29544)(6.8,2.41923)(6.7,2.56286)(6.6,2.72839)(6.5,2.91763)(6.4,3.13197)(6.3,3.37217)(6.2,3.63821)(6.1,3.92921)(6.,4.24332)(5.9,4.57786)(5.8,4.92937)(5.7,5.29392)(5.6,5.66735)(5.5,6.04556)(5.4,6.42478)(5.3,6.80179)(5.2,7.174)(5.1,7.53956)(5.,7.89729)(4.9,8.24665)(4.8,8.5876)(4.7,8.92053)(4.6,9.24613)(4.5,9.56529)(4.4,9.87903)(4.3,10.1884)(4.2,10.4946)(4.1,10.7985)(4.,11.1013)(3.9,11.4038)(3.8,11.707)(3.7,12.0116)(3.6,12.3184)(3.5,12.6281)(3.4,12.9412)(3.3,13.2583)(3.2,13.5799)(3.1,13.9064)(3.,14.2384)(2.9,14.576)(2.8,14.9198)(2.7,15.27)(2.6,15.6269)(2.5,15.9907)(2.4,16.3619)(2.3,16.7406)(2.2,17.1271)(2.1,17.5215)(2.,17.9242)(1.9,18.3354)(1.8,18.7552)(1.7,19.1839)(1.6,19.6218)(1.5,20.0689)(1.4,20.5256)(1.3,20.992)(1.2,21.4683)(1.1,21.9548)(1.,22.4517)
};
	\end{axis}
	\end{tikzpicture}
	\begin{tikzpicture}
	\begin{axis}[width=0.29\textwidth, ymax=150, xlabel={$|F|$ [pN]},
	xlabel near ticks, ylabel near ticks, ylabel shift = -4 pt] 
	\addplot[color=black, no markers,style={thick}]  coordinates{(10,0.828573 )(99/10,0.837597 )(49/5,0.845885 )(97/10,0.853231 )(48/5,0.859389 )(19/2,0.864061 )(47/5,0.866891 )(93/10,0.867457 )(46/5,0.865261 )(91/10,0.859725 )(9,0.850184 )(89/10,0.835891 )(44/5,0.816026 )(87/10,0.789715 )(43/5,0.756077 )(17/2,0.714284 )(42/5,0.663658 )(83/10,0.603808 )(41/5,0.534808 )(81/10,0.45741 )(8,0.373305 )(79/10,0.28539 )(39/5,0.198033 )(77/10,0.117289 )(38/5,0.0510366 )(15/2,0.0089941 )(37/5,0.00260056 )(73/10,0.0447654 )(36/5,0.149508 )(71/10,0.33153 )(7,0.605766 )(69/10,0.986961 )(34/5,1.4893 )(67/10,2.12609 )(33/5,2.90959 )(13/2,3.8508 )(32/5,4.95942 )(63/10,6.24368 )(31/5,7.71021 )(61/10,9.36365 )(6,11.2062 )(59/10,13.2366 )(29/5,15.4497 )(57/10,17.835 )(28/5,20.3764 )(11/2,23.052 )(27/5,25.8344 )(53/10,28.6919 )(26/5,31.5906 )(51/10,34.4961 )(5,37.376 )(49/10,40.2017 )(24/5,42.9497 )(47/10,45.6025 )(23/5,48.1485 )(9/2,50.5818 )(22/5,52.9013 )(43/10,55.1098 )(21/5,57.2129 )(41/10,59.2183 )(4,61.1347 )(39/10,62.9715 )(19/5,64.738 )(37/10,66.4432 )(18/5,68.0955 )(7/2,69.7027 )(17/5,71.2717 )(33/10,72.8086 )(16/5,74.319 )(31/10,75.8074 )(3,77.278 )(29/10,78.7341 )(14/5,80.1788 )(27/10,81.6143 )(13/5,83.0428 )(5/2,84.4658 )(12/5,85.8848 )(23/10,87.3008 )(11/5,88.7146 )(21/10,90.127 )(2,91.5384 )(19/10,92.9492 )(9/5,94.3595 )(17/10,95.7697 )(8/5,97.1796 )(3/2,98.5893 )(7/5,99.9987 )(13/10,101.408 )(6/5,102.816 )(11/10,104.224 )(1,105.631)};
	\addplot[color=webgreen, no markers,style={thick},dotted] coordinates{(10,101.542 )(99/10,102.334 )(49/5,103.131 )(97/10,103.931 )(48/5,104.736 )(19/2,105.543 )(47/5,106.353 )(93/10,107.165 )(46/5,107.978 )(91/10,108.79 )(9,109.6 )(89/10,110.406 )(44/5,111.206 )(87/10,111.997 )(43/5,112.777 )(17/2,113.541 )(42/5,114.285 )(83/10,115.003 )(41/5,115.688 )(81/10,116.334 )(8,116.93 )(79/10,117.465 )(39/5,117.928 )(77/10,118.302 )(38/5,118.571 )(15/2,118.715 )(37/5,118.714 )(73/10,118.542 )(36/5,118.175 )(71/10,117.586 )(7,116.75 )(69/10,115.64 )(34/5,114.238 )(67/10,112.528 )(33/5,110.506 )(13/2,108.176 )(32/5,105.559 )(63/10,102.69 )(31/5,99.6171 )(61/10,96.4043 )(6,93.1235 )(59/10,89.8521 )(29/5,86.6669 )(57/10,83.639 )(28/5,80.8291 )(11/2,78.2847 )(27/5,76.0386 )(53/10,74.1089 )(26/5,72.5006 )(51/10,71.2079 )(5,70.2167 )(49/10,69.5071 )(24/5,69.0558 )(47/10,68.8379 )(23/5,68.8285 )(9/2,69.0037 )(22/5,69.3409 )(43/10,69.8199 )(21/5,70.4224 )(41/10,71.1324 )(4,71.9358 )(39/10,72.8206 )(19/5,73.7764 )(37/10,74.7945 )(18/5,75.8674 )(7/2,76.989 )(17/5,78.1539 )(33/10,79.3578 )(16/5,80.5972 )(31/10,81.8688 )(3,83.1704 )(29/10,84.4997 )(14/5,85.8552 )(27/10,87.2353 )(13/5,88.6391 )(5/2,90.0655 )(12/5,91.5138 )(23/10,92.9834 )(11/5,94.4738 )(21/10,95.9846 )(2,97.5156 )(19/10,99.0665 )(9/5,100.637 )(17/10,102.227 )(8/5,103.837 )(3/2,105.466 )(7/5,107.114 )(13/10,108.782 )(6/5,110.47 )(11/10,112.176 )(1,113.902)
	};
	\addplot[color=red, no markers,style={thick},dash pattern={on 4pt off 1pt on 2pt off 3pt}] coordinates{(10,67.3297 )(99/10,67.8242 )(49/5,68.3198 )(97/10,68.8163 )(48/5,69.3132 )(19/2,69.81 )(47/5,70.3061 )(93/10,70.8008 )(46/5,71.2934 )(91/10,71.7828 )(9,72.268 )(89/10,72.7475 )(44/5,73.2198 )(87/10,73.6831 )(43/5,74.1353 )(17/2,74.5739 )(42/5,74.9961 )(83/10,75.3988 )(41/5,75.7781 )(81/10,76.1301 )(8,76.4502 )(79/10,76.7334 )(39/5,76.9743 )(77/10,77.1671 )(38/5,77.3063 )(15/2,77.3862 )(37/5,77.4018 )(73/10,77.3491 )(36/5,77.2259 )(71/10,77.0326 )(7,76.7733 )(69/10,76.4565 )(34/5,76.0967 )(67/10,75.7153 )(33/5,75.3409 )(13/2,75.01 )(32/5,74.766 )(63/10,74.6579 )(31/5,74.738 )(61/10,75.0583 )(6,75.6676 )(59/10,76.6076 )(29/5,77.9104 )(57/10,79.5964 )(28/5,81.6742 )(11/2,84.1411 )(27/5,86.9848 )(53/10,90.1862 )(26/5,93.7211 )(51/10,97.5634 )(5,101.687 )(49/10,106.066 )(24/5,110.678 )(47/10,115.502 )(23/5,120.523 )(9/2,125.727 )(22/5,131.101 )(43/10,136.639 )(21/5,142.334 )(41/10,148.182 )(4,154.18 )(39/10,160.328 )(19/5,166.625 )(37/10,173.071 )(18/5,179.669 )(7/2,186.419 )(17/5,193.325 )(33/10,200.388 )(16/5,207.611 )(31/10,214.998 )(3,222.55 )(29/10,230.271 )(14/5,238.164 )(27/10,246.233 )(13/5,254.479 )(5/2,262.907 )(12/5,271.518 )(23/10,280.317 )(11/5,289.307 )(21/10,298.49 )(2,307.869 )(19/10,317.447 )(9/5,327.228 )(17/10,337.214 )(8/5,347.409 )(3/2,357.814 )(7/5,368.434 )(13/10,379.271 )(6/5,390.328 )(11/10,401.607 )(1,413.113)};
	   \addplot[color=blue, no markers,style={thick}, dashed]  coordinates{	
	   (10.,14.6248)(9.9,14.7385)(9.8,14.8527)(9.7,14.9675)(9.6,15.0826)(9.5,15.198)(9.4,15.3135)(9.3,15.429)(9.2,15.5443)(9.1,15.6591)(9.,15.7732)(8.9,15.886)(8.8,15.9973)(8.7,16.1064)(8.6,16.2127)(8.5,16.3153)(8.4,16.4427)(8.3,16.6082)(8.2,16.7841)(8.1,16.9726)(8.,17.1764)(7.9,17.3987)(7.8,17.6434)(7.7,17.9152)(7.6,18.2195)(7.5,18.5631)(7.4,18.9534)(7.3,19.3991)(7.2,19.91)(7.1,20.4969)(7.,21.1714)(6.9,21.9453)(6.8,22.8305)(6.7,23.8376)(6.6,24.9755)(6.5,26.2503)(6.4,27.6637)(6.3,29.213)(6.2,30.8898)(6.1,32.6804)(6.,34.5664)(5.9,36.5254)(5.8,38.5331)(5.7,40.5646)(5.6,42.5966)(5.5,44.6084)(5.4,46.5831)(5.3,48.5084)(5.2,50.3757)(5.1,52.1809)(5.,53.9229)(4.9,55.6034)(4.8,57.226)(4.7,58.7958)(4.6,60.3184)(4.5,61.8)(4.4,63.2466)(4.3,64.6641)(4.2,66.0581)(4.1,67.4337)(4.,68.7956)(3.9,70.1479)(3.8,71.4943)(3.7,72.8381)(3.6,74.1821)(3.5,75.5288)(3.4,76.8803)(3.3,78.2386)(3.2,79.6052)(3.1,80.9815)(3.,82.3688)(2.9,83.768)(2.8,85.1801)(2.7,86.6058)(2.6,88.0459)(2.5,89.5009)(2.4,90.9713)(2.3,92.4576)(2.2,93.9601)(2.1,95.4792)(2.,97.0153)(1.9,98.5684)(1.8,100.139)(1.7,101.727)(1.6,103.333)(1.5,104.957)(1.4,106.6)(1.3,108.26)(1.2,109.939)(1.1,111.637)(1.,113.353)
	   };
	\end{axis}
	\end{tikzpicture}
	\begin{tikzpicture}
	\begin{axis}[width=0.29\textwidth, ymax=400, xlabel={$|F|$ [pN]},
	xlabel near ticks, ylabel near ticks, ylabel shift = -4 pt] 
	\addplot[color=black, no markers,style={thick}]  coordinates{
		(10,0.771419 )(99/10,0.767929 )(49/5,0.761154 )(97/10,0.750458 )(48/5,0.735137 )(19/2,0.714429 )(47/5,0.68754 )(93/10,0.653697 )(46/5,0.612214 )(91/10,0.562597 )(9,0.504686 )(89/10,0.438835 )(44/5,0.366126 )(87/10,0.288617 )(43/5,0.209601 )(17/2,0.13385 )(42/5,0.0678254 )(83/10,0.0198226 )(41/5,0.0000399943 )(81/10,0.0205716 )(8,0.0953421 )(79/10,0.240013 )(39/5,0.471889 )(77/10,0.809838 )(38/5,1.27423 )(15/2,1.88683 )(37/5,2.67065 )(73/10,3.6496 )(36/5,4.84798 )(71/10,6.28961 )(7,7.99677 )(69/10,9.98872 )(34/5,12.2801 )(67/10,14.8791 )(33/5,17.7855 )(13/2,20.9891 )(32/5,24.4684 )(63/10,28.1893 )(31/5,32.106 )(61/10,36.1615 )(6,40.2907 )(59/10,44.4241 )(29/5,48.492 )(57/10,52.4301 )(28/5,56.1834 )(11/2,59.7091 )(27/5,62.9786 )(53/10,65.9767 )(26/5,68.701 )(51/10,71.159 )(5,73.3661 )(49/10,75.3423 )(24/5,77.1106 )(47/10,78.6951 )(23/5,80.1191 )(9/2,81.4051 )(22/5,82.5733 )(43/10,83.6421 )(21/5,84.6276 )(41/10,85.544 )(4,86.4031 )(39/10,87.2154 )(19/5,87.9895 )(37/10,88.7328 )(18/5,89.4515 )(7/2,90.1505 )(17/5,90.8343 )(33/10,91.5064 )(16/5,92.1698 )(31/10,92.8269 )(3,93.4797 )(29/10,94.1299 )(14/5,94.779 )(27/10,95.4281 )(13/5,96.0781 )(5/2,96.7299 )(12/5,97.3842 )(23/10,98.0414 )(11/5,98.7021 )(21/10,99.3667 )(2,100.035 )(19/10,100.709 )(9/5,101.387 )(17/10,102.07 )(8/5,102.758 )(3/2,103.451 )(7/5,104.15 )(13/10,104.854 )(6/5,105.564 )(11/10,106.28 )(1,107.001)};
	\addplot[color=webgreen, no markers,style={thick},dotted] coordinates{
		(10,101.566 )(99/10,102.363 )(49/5,103.166 )(97/10,103.974 )(48/5,104.788 )(19/2,105.606 )(47/5,106.43 )(93/10,107.258 )(46/5,108.09 )(91/10,108.926 )(9,109.764 )(89/10,110.605 )(44/5,111.448 )(87/10,112.29 )(43/5,113.132 )(17/2,113.971 )(42/5,114.805 )(83/10,115.633 )(41/5,116.451 )(81/10,117.257 )(8,118.046 )(79/10,118.815 )(39/5,119.559 )(77/10,120.272 )(38/5,120.949 )(15/2,121.582 )(37/5,122.167 )(73/10,122.695 )(36/5,123.162 )(71/10,123.563 )(7,123.894 )(69/10,124.155 )(34/5,124.35 )(67/10,124.484 )(33/5,124.569 )(13/2,124.62 )(32/5,124.654 )(63/10,124.693 )(31/5,124.758 )(61/10,124.869 )(6,125.045 )(59/10,125.3 )(29/5,125.644 )(57/10,126.084 )(28/5,126.622 )(11/2,127.255 )(27/5,127.981 )(53/10,128.792 )(26/5,129.682 )(51/10,130.643 )(5,131.668 )(49/10,132.749 )(24/5,133.88 )(47/10,135.054 )(23/5,136.267 )(9/2,137.513 )(22/5,138.79 )(43/10,140.093 )(21/5,141.42 )(41/10,142.768 )(4,144.136 )(39/10,145.521 )(19/5,146.924 )(37/10,148.341 )(18/5,149.774 )(7/2,151.22 )(17/5,152.68 )(33/10,154.153 )(16/5,155.638 )(31/10,157.136 )(3,158.646 )(29/10,160.168 )(14/5,161.702 )(27/10,163.248 )(13/5,164.806 )(5/2,166.375 )(12/5,167.957 )(23/10,169.55 )(11/5,171.156 )(21/10,172.773 )(2,174.402 )(19/10,176.043 )(9/5,177.697 )(17/10,179.363 )(8/5,181.041 )(3/2,182.731 )(7/5,184.435 )(13/10,186.15 )(6/5,187.879 )(11/10,189.62 )(1,191.375)};
	\addplot[color=red, no markers,style={thick},dash pattern={on 4pt off 1pt on 2pt off 3pt}] coordinates{
		(10,305.64 )(99/10,307.977 )(49/5,310.322 )(97/10,312.673 )(48/5,315.028 )(19/2,317.384 )(47/5,319.738 )(93/10,322.085 )(46/5,324.421 )(91/10,326.739 )(9,329.032 )(89/10,331.292 )(44/5,333.507 )(87/10,335.666 )(43/5,337.753 )(17/2,339.751 )(42/5,341.639 )(83/10,343.393 )(41/5,344.985 )(81/10,346.381 )(8,347.544 )(79/10,348.432 )(39/5,348.997 )(77/10,349.188 )(38/5,348.95 )(15/2,348.226 )(37/5,346.959 )(73/10,345.101 )(36/5,342.61 )(71/10,339.461 )(7,335.653 )(69/10,331.216 )(34/5,326.216 )(67/10,320.759 )(33/5,314.996 )(13/2,309.113 )(32/5,303.326 )(63/10,297.863 )(31/5,292.951 )(61/10,288.797 )(6,285.573 )(59/10,283.405 )(29/5,282.372 )(57/10,282.504 )(28/5,283.787 )(11/2,286.176 )(27/5,289.601 )(53/10,293.976 )(26/5,299.211 )(51/10,305.214 )(5,311.897 )(49/10,319.179 )(24/5,326.988 )(47/10,335.261 )(23/5,343.944 )(9/2,352.991 )(22/5,362.364 )(43/10,372.03 )(21/5,381.965 )(41/10,392.147 )(4,402.559 )(39/10,413.189 )(19/5,424.026 )(37/10,435.062 )(18/5,446.291 )(7/2,457.709 )(17/5,469.311 )(33/10,481.097 )(16/5,493.064 )(31/10,505.212 )(3,517.54 )(29/10,530.049 )(14/5,542.74 )(27/10,555.613 )(13/5,568.67 )(5/2,581.911 )(12/5,595.338 )(23/10,608.952 )(11/5,622.756 )(21/10,636.751 )(2,650.938 )(19/10,665.32 )(9/5,679.898 )(17/10,694.674 )(8/5,709.65 )(3/2,724.828 )(7/5,740.211 )(13/10,755.799 )(6/5,771.596 )(11/10,787.603 )(1,803.822)};		
	   \addplot[color=blue, no markers,style={thick}, dashed]  coordinates{	
(10.,63.6755)(9.9,64.1678)(9.8,64.6619)(9.7,65.1572)(9.6,65.6532)(9.5,66.149)(9.4,66.6438)(9.3,67.1364)(9.2,67.6299)(9.1,68.1739)(9.,68.7253)(8.9,69.2849)(8.8,69.8537)(8.7,70.4328)(8.6,71.0238)(8.5,71.6284)(8.4,72.2488)(8.3,72.8877)(8.2,73.5484)(8.1,74.2348)(8.,74.9516)(7.9,75.7043)(7.8,76.4996)(7.7,77.345)(7.6,78.2494)(7.5,79.2227)(7.4,80.2758)(7.3,81.4203)(7.2,82.668)(7.1,84.0306)(7.,85.5184)(6.9,87.1392)(6.8,88.8977)(6.7,90.7937)(6.6,92.8214)(6.5,94.969)(6.4,97.2186)(6.3,99.5474)(6.2,101.929)(6.1,104.335)(6.,106.739)(5.9,109.115)(5.8,111.443)(5.7,113.707)(5.6,115.897)(5.5,118.008)(5.4,120.037)(5.3,121.988)(5.2,123.864)(5.1,125.672)(5.,127.419)(4.9,129.112)(4.8,130.759)(4.7,132.367)(4.6,133.942)(4.5,135.49)(4.4,137.017)(4.3,138.528)(4.2,140.025)(4.1,141.514)(4.,142.996)(3.9,144.476)(3.8,145.953)(3.7,147.432)(3.6,148.913)(3.5,150.398)(3.4,151.887)(3.3,153.383)(3.2,154.885)(3.1,156.395)(3.,157.913)(2.9,159.439)(2.8,160.975)(2.7,162.52)(2.6,164.075)(2.5,165.64)(2.4,167.215)(2.3,168.801)(2.2,170.398)(2.1,172.006)(2.,173.625)(1.9,175.256)(1.8,176.898)(1.7,178.552)(1.6,180.218)(1.5,181.896)(1.4,183.586)(1.3,185.288)(1.2,187.003)(1.1,188.73)(1.,190.47)	   
	   };	
	\end{axis}
	\end{tikzpicture}
	\caption{Precision $\mathfrak{p}$ (solid) for the kinesin displacement along the microtubule. Comparison with the absolute current bound \eqref{abscurr} (dashed), the activity bound \eqref{activity} (dotted), and the entropy production bound \eqref{sigma} (dash-dotted). From left to right $\text{[ATP]}=1, 10, 10^2, 10^3 \, \mu$M.}
	\label{fig:kin}
\end{figure*}


A slightly tighter bound than \eqref{eq:vdb} (whose implicit expression appears in \cite{tim19}) follows from replacing $d$ in \eqref{eq:general} by an upper bound given in terms of $\mean{\sigma}$ (see SI), leading to  the novel asymptotic expression 
\begin{align}\label{bestp}
\mathfrak{p}_\text{max} \leq  e^{\langle \sigma  \rangle}/4 \quad \text{ for } \mean{\sigma} \gg 1.
\end{align} 
Remarkably, this is the tightest bound obtainable from the sole knowledge of $\mean{\sigma}$. This is proved in SI by finding a specific system $\Omega$ and a specific observable on $\Omega$  that meets the bound, for every given value of $\mean{\sigma}$.

\emph{The Periodic Uncertainty Relation---} In many cases it is relevant to decompose a path $\omega$ becomes the concatenation of paths $\omega_0, \omega_1, \ldots, \omega_{N-1}$ taking place on $N$ time intervals of duration $\Delta t$. In this way the space $\Omega$ of paths factors as a Cartesian product  $\Omega_0 \times \Omega_1 \times  \ldots \Omega_{N-1}$. 
Here we study the most common case of interest where every path $\omega_i$ is a (discrete or continuous) sequence of states and transitions in a Markov process, and where the  the sequence $\omega_0, \omega_1, \ldots, \omega_{N-1}$ is stationary (periodicity assumption). This is typically sufficient to model overdamped Markov processes. We also consider the legitimate observables on $\omega$ as those observables $f$ that decompose as a sum $f(\omega)=f_0 (\omega_0)+ \ldots + f_{N-1}(\omega_{N-1})$, where each $f_i$ is time-antisymmetric: $f_i(\omega_i)=-f_i(\overline{\omega}_i)$. In this case we find that the precision available over any number $N$ of periods is bounded above by the precision available over a single period, namely,
\begin{equation}\label{eq:periodicallN}
\mathfrak{p}_\text{max}(\Omega)/ N  \leq \mathfrak{p}_\text{max}(\Omega_i),
\end{equation}
a theorem (proved in SI) that we call the Periodic Uncertainty Relation for time-antisymmetric observables on overdamped Markov processes. In particular, applying \eqref{eq:vdb} to a single period, we find that the precision of $N$ periods is bounded by $(e^\mean{\sigma}-1)/2$, where $\langle \sigma \rangle$ is now the Kullback-Leibler divergence over a single interval. This is valid for arbitrary protocols (being understood that $\mean{\sigma}$ is not necessarily the entropy production). This includes in particular the result in \cite{pro17}, which was proved originally by large deviation techniques in the limit $N \to \infty$,  for overdamped systems under time-symmetric protocols. Our result is a special case of a more general Periodic Uncertainty Relation, stated and proved in the SI, which holds for more general families of legitimate observables. 

In the case of stationary (jump or diffusive) processes over a total time interval $[0,t_{\text{f}}]$, the period is infinitesimal, $\Delta t = t_{\text{f}}/N \to 0$, and so is  $\mean{\sigma}$. Then \eqref{eq:vdb} combined with \eqref{eq:periodicallN} reduces to 
\begin{align}\label{sigma}
\mathfrak{p}_\text{max}/{t_f} \leq \mean{\sigma}/2,
\end{align}
so that we recover the entropy bound for arbitrary time intervals, previously proved with information-theoretic means \cite{dec18}. Beyond recovering these results with a unified method, we can derive far sharper bounds. In particular, for a stationary continuous-time Markov process, precision and normalized precision over an infinitesimal time interval $\Delta t$ coincide. So, in view of \eqref{eq:periodicallN}, the precision over a time interval $t_{\text{f}}=N \Delta t$ is bounded by 
\begin{align}\label{p2p}
\frac{\mathfrak{p}_\text{max}}{t_{\text{f}}} \leq  \frac{1}{\Delta t}\sum_{e} \frac{(p_{e}-p_{\overline{e}})^2}{p_{e}+p_{\overline{e}}}
\end{align}
where $p_{e}$ is the probability of a transition along a path $e$  relating a source state $s(e)$ to a target state $t(e)$ over an infinitesimal time interval $\Delta t$. The current is defined as $j_e:=p_e/\Delta t$.   In a finite state jump process, $e$ is a transition between two different states, and $j_{e}$ factors as $p_{s(e)} w_{e}$ for stationary state probability $p_{s(e)}$ and jumping rate $w_{e}$.
We know that \eqref{p2p}  can be relaxed to $ \mathfrak{p}_\text{max}/t_{\text{f}} \leq (d/\Delta t)  \tanh ( \langle \sigma \rangle/2d)$ with  $d=\frac{1}{2}\sum_{e}|p_{e}-p_{\overline{e}}|$. We obtain in particular the simple and novel bound,
\begin{align}\label{abscurr}
\frac{\mathfrak{p}_\text{max}}{t_{\text{f}}} \leq \frac{d}{ \Delta t} = \frac{1}{2}\sum_{e} | j_e- j_{\overline{e}}|
\end{align}
which we call the \emph{absolute current bound}, valid for all stationary Markov processes.

In the case of finite state jump processes, it is evidently tighter than the activity bound,  
\begin{align}\label{activity}
\frac{\mathfrak{p}_\text{max}}{t_{\text{f}}} \leq  \frac{1}{2}\sum_e ( j_e + j_{\overline{e}}).
\end{align}
This last bound applies to all time-summed observables taking non-zero values only on the transitions (thus zero values on the constant paths), without any request of time-antisymmetry  \cite{dit18}. The activity bound turns out to be another avatar of the Periodic Uncertainty Relation, where the bound can be derived on an infinitesimal interval and then extended to arbitrary times (see SI).

\emph{Example---}We illustrate the different bounds for stationary Markovian dynamics on a benchmark example \cite{dit18} which provides a minimal model for the molecular motor kinesin moving under load along a microtubule. Kinesin is either in a low energy state (1) with both heads on the microtubule or in a high energy state (2) with only one head attached. Transitions from state 1 to state 2 happen with or without ATP consumption, and cause both forward and backward motion along the microtubule (with half step size $\ell \simeq 4 \,\text{nm} $). Each of these four transitions $e=1,\dots, 4$ out of each state $x=1,2$ (making eight possible transitions) has an associated rate $w_{xe}$, function of the ATP concentration, [ATP], and of the external loading force $F<0$ (see SI).
In Fig.~\eqref{fig:kin} we plot $\mathfrak{p}$ for the displacement, and the various bounds. We see that our simpler novel bound \eqref{abscurr} outperforms the activity bound and entropy production bounds.

\emph{Discussion}---The general approach introduced in this Letter solely exploits the properties of the Hilbert space of observables and the presence of a (broken) involutive symmetry. Therefore, it is not restricted to trajectories of random systems endowed with some notion of time-reversal symmetry. Rather, $\Omega$ can be, e.g., the configuration space of a classical or quantum system and the involution may be parity, charge conjugation, spin reversal, etc. (see SI for an example of an Ising system). We leave for the future the application to quantum systems and spontaneously broken symmetries.

 \emph{Acknowledgments}--- M.~E. thanks the European Research Council (project NanoThermo ERC-2015-COG agreement no. 681456).  M.~E. and J-C.~D. thank the FNR INTER mobility program.

\newpage 
\begin{widetext}
\section{Supplemental informations}

\subsection{Best bound based on $\mean{\sigma}$ only}

A slightly tighter bound than \eqref{eq:vdb} follows by bounding $d$ in terms of $\mean{\sigma}$. Indeed, for the legitimate observable $f=\text{sgn}(p-\overline{p})$, the normalized precision \eqref{np} is $d^2$ and \eqref{eq:general} becomes 
\begin{align}\label{d}
d^2 \leq d \tanh \frac{\langle\sigma\rangle}{2d},
\end{align}
or
\begin{align}\label{d2}
2 d\, \text{atanh} \, d \leq \langle\sigma\rangle,
\end{align}
which allows to find the bound $d \leq d^*(\mean{\sigma})$ where the r.h.s. is defined by
\begin{align} \label{dstar}
2 d^*\, \text{atanh} \, d^*= \langle\sigma\rangle.
\end{align}
 Injecting this bound on $d$ into  \eqref{eq:general}, we obtain 
\begin{align}\label{implicit}
\mathfrak{np}_\text{max} \leq \ d^* \tanh \frac{\langle\sigma\rangle}{2 d^*}=d^*(\mean{\sigma})^2,
\end{align}
a tighter bound than  \eqref{eq:vdb}, which was obtained from the trivial bound $d\leq 1$ --- see figure \ref{fig:ratio} for  a comparison of the two bounds.

Remarkably, this is the tightest bound obtainable from the sole knowledge of $\mean{\sigma}$ as the following argument proves. We split $\Omega$ into $\Omega_0$ and $\overline{\Omega}_0$. On both parts we take a uniform probability distribution, so that the total probability of  $\Omega_0$ is $p_0 \geq 1/2$, chosen to satisfy $(2p_0-1)\ln \frac{p_0}{1-p_0}=\mean{\sigma}$. One sees that the total variation distance is precisely $d=2p_0-1$, and the bound is matched with equality. 

The asymptotic expression \eqref{bestp} is obtained  for $\mean{\sigma} \gg 1$, or equivalently $1-d^* \ll 1$. 
We expand $d^*=\tanh \frac{\mean{\sigma}}{2d^*} \approx 1-2e^{-\frac{\mean{\sigma}}{d^*}} \approx 1-2e^{-\mean{\sigma}}$. The latter step stems from $d^* \approx 1$ but  requires some care in the error analysis, in particular it requires to show that $(1-d^*)\mean{\sigma} \ll 1$.


Plugging $d^*$ into \eqref{implicit} and using the relation $\mathfrak{p}=\mathfrak{np}/(1-\mathfrak{np})$, we obtain \eqref{bestp}, confirming  the numerical observation in figure \ref{fig:ratio}.

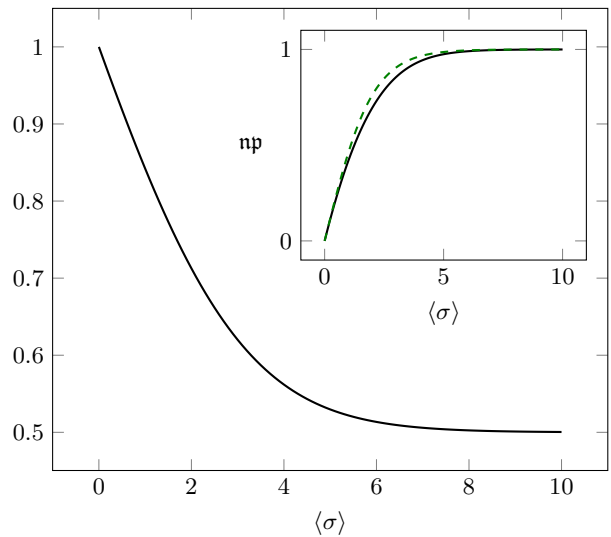
\begin{figure}[h]
	\centering
	\begin{tikzpicture}
	\begin{axis}[width=0.5\textwidth, xlabel=${\mean{\sigma}}$,
	xlabel near ticks, ylabel near ticks, ylabel shift = -4 pt] 
	\addplot[color=black, no markers,style={thick}] coordinates{
	(0.0001,0.999983)(0.1001,0.983376)(0.2001,0.9669)(0.3001,0.950576)(0.4001,0.934423)(0.5001,0.918462)(0.6001,0.90271)(0.7001,0.887186)(0.8001,0.871907)(0.9001,0.85689)(1.0001,0.84215)(1.1001,0.827701)(1.2001,0.813558)(1.3001,0.799734)(1.4001,0.786238)(1.5001,0.773083)(1.6001,0.760276)(1.7001,0.747827)(1.8001,0.735741)(1.9001,0.724024)(2.0001,0.712682)(2.1001,0.701716)(2.2001,0.69113)(2.3001,0.680924)(2.4001,0.671097)(2.5001,0.66165)(2.6001,0.652579)(2.7001,0.643882)(2.8001,0.635554)(2.9001,0.62759)(3.0001,0.619985)(3.1001,0.612732)(3.2001,0.605824)(3.3001,0.599252)(3.4001,0.593009)(3.5001,0.587085)(3.6001,0.581471)(3.7001,0.576158)(3.8001,0.571135)(3.9001,0.566392)(4.0001,0.561919)(4.1001,0.557704)(4.2001,0.553738)(4.3001,0.55001)(4.4001,0.546509)(4.5001,0.543225)(4.6001,0.540147)(4.7001,0.537265)(4.8001,0.534569)(4.9001,0.53205)(5.0001,0.529698)(5.1001,0.527504)(5.2001,0.525458)(5.3001,0.523553)(5.4001,0.521779)(5.5001,0.52013)(5.6001,0.518597)(5.7001,0.517174)(5.8001,0.515852)(5.9001,0.514627)(6.0001,0.513491)(6.1001,0.512438)(6.2001,0.511463)(6.3001,0.510561)(6.4001,0.509727)(6.5001,0.508956)(6.6001,0.508243)(6.7001,0.507584)(6.8001,0.506977)(6.9001,0.506416)(7.0001,0.505898)(7.1001,0.505421)(7.2001,0.504981)(7.3001,0.504576)(7.4001,0.504203)(7.5001,0.503859)(7.6001,0.503542)(7.7001,0.503251)(7.8001,0.502983)(7.9001,0.502737)(8.0001,0.50251)(8.1001,0.502302)(8.2001,0.502111)(8.3001,0.501935)(8.4001,0.501774)(8.5001,0.501625)(8.6001,0.501489)(8.7001,0.501364)(8.8001,0.50125)(8.9001,0.501144)(9.0001,0.501048)(9.1001,0.500959)(9.2001,0.500878)(9.3001,0.500804)(9.4001,0.500736)(9.5001,0.500673)(9.6001,0.500616)(9.7001,0.500563)(9.8001,0.500515)(9.9001,0.500471)(10.0001,0.500431)
};	
\end{axis}
\begin{axis}[anchor=south west, xshift=3.3cm, yshift=2.8cm,width=0.3\textwidth, ylabel=$\mathfrak{np}$, xlabel=$\mean{\sigma}$, ytick={0,1},ylabel near ticks, yticklabel pos=left, xlabel near ticks,ylabel style={rotate=-90}]
\addplot[color=black, no markers,style={thick},solid] coordinates{
(0.,0.)(0.1,0.0491695)(0.2,0.0966894)(0.3,0.142578)(0.4,0.186853)(0.5,0.229535)(0.6,0.270645)(0.7,0.310203)(0.8,0.348232)(0.9,0.384756)(1.,0.419798)(1.1,0.453385)(1.2,0.485542)(1.3,0.516296)(1.4,0.545677)(1.5,0.573713)(1.6,0.600435)(1.7,0.625874)(1.8,0.650062)(1.9,0.673031)(2.,0.694817)(2.1,0.715452)(2.2,0.734973)(2.3,0.753416)(2.4,0.770816)(2.5,0.787211)(2.6,0.802639)(2.7,0.817135)(2.8,0.830738)(2.9,0.843487)(3.,0.855417)(3.1,0.866567)(3.2,0.876974)(3.3,0.886674)(3.4,0.895704)(3.5,0.904098)(3.6,0.911893)(3.7,0.91912)(3.8,0.925815)(3.9,0.932007)(4.,0.93773)(4.1,0.943011)(4.2,0.94788)(4.3,0.952365)(4.4,0.956491)(4.5,0.960283)(4.6,0.963766)(4.7,0.966961)(4.8,0.969889)(4.9,0.972572)(5.,0.975026)(5.1,0.977271)(5.2,0.979323)(5.3,0.981197)(5.4,0.982907)(5.5,0.984466)(5.6,0.985888)(5.7,0.987184)(5.8,0.988363)(5.9,0.989437)(6.,0.990414)(6.1,0.991303)(6.2,0.992111)(6.3,0.992845)(6.4,0.993512)(6.5,0.994118)(6.6,0.994668)(6.7,0.995167)(6.8,0.995621)(6.9,0.996032)(7.,0.996405)(7.1,0.996743)(7.2,0.99705)(7.3,0.997328)(7.4,0.99758)(7.5,0.997808)(7.6,0.998015)(7.7,0.998203)(7.8,0.998373)(7.9,0.998527)(8.,0.998666)(8.1,0.998793)(8.2,0.998907)(8.3,0.99901)(8.4,0.999104)(8.5,0.999189)(8.6,0.999266)(8.7,0.999336)(8.8,0.999399)(8.9,0.999456)(9.,0.999508)(9.1,0.999554)(9.2,0.999597)(9.3,0.999635)(9.4,0.99967)(9.5,0.999701)(9.6,0.999729)(9.7,0.999755)(9.8,0.999778)(9.9,0.9998)(10.,0.999819)
};
\addplot[color=webgreen, no markers,style={thick},dashed] coordinates{
(0.,0.)(0.1,0.0499584)(0.2,0.099668)(0.3,0.148885)(0.4,0.197375)(0.5,0.244919)(0.6,0.291313)(0.7,0.336376)(0.8,0.379949)(0.9,0.421899)(1.,0.462117)(1.1,0.50052)(1.2,0.53705)(1.3,0.57167)(1.4,0.604368)(1.5,0.635149)(1.6,0.664037)(1.7,0.691069)(1.8,0.716298)(1.9,0.739783)(2.,0.761594)(2.1,0.781806)(2.2,0.800499)(2.3,0.817754)(2.4,0.833655)(2.5,0.848284)(2.6,0.861723)(2.7,0.874053)(2.8,0.885352)(2.9,0.895693)(3.,0.905148)(3.1,0.913785)(3.2,0.921669)(3.3,0.928858)(3.4,0.935409)(3.5,0.941376)(3.6,0.946806)(3.7,0.951746)(3.8,0.956237)(3.9,0.960319)(4.,0.964028)(4.1,0.967395)(4.2,0.970452)(4.3,0.973226)(4.4,0.975743)(4.5,0.978026)(4.6,0.980096)(4.7,0.981973)(4.8,0.983675)(4.9,0.985217)(5.,0.986614)(5.1,0.98788)(5.2,0.989027)(5.3,0.990066)(5.4,0.991007)(5.5,0.99186)(5.6,0.992632)(5.7,0.99333)(5.8,0.993963)(5.9,0.994536)(6.,0.995055)(6.1,0.995524)(6.2,0.995949)(6.3,0.996334)(6.4,0.996682)(6.5,0.996998)(6.6,0.997283)(6.7,0.997541)(6.8,0.997775)(6.9,0.997986)(7.,0.998178)(7.1,0.998351)(7.2,0.998508)(7.3,0.99865)(7.4,0.998778)(7.5,0.998894)(7.6,0.999)(7.7,0.999095)(7.8,0.999181)(7.9,0.999259)(8.,0.999329)(8.1,0.999393)(8.2,0.999451)(8.3,0.999503)(8.4,0.99955)(8.5,0.999593)(8.6,0.999632)(8.7,0.999667)(8.8,0.999699)(8.9,0.999727)(9.,0.999753)(9.1,0.999777)(9.2,0.999798)(9.3,0.999817)(9.4,0.999835)(9.5,0.99985)(9.6,0.999865)(9.7,0.999877)(9.8,0.999889)(9.9,0.9999)(10.,0.999909)
};
\end{axis}
	\end{tikzpicture}
	\caption[]{Ratio between the best bound on precision obtained from the knowledge of $\mean{\sigma}$ alone, \eqref{implicit}, and the bound \eqref{eq:vdb}. {\em Inset}: the bound \eqref{eq:general} with $d=d^*(\mean{\sigma})$ given by \eqref{dstar}, i.e. \eqref{implicit} (solid), and with $d=1$ (dashed).}
	\label{fig:ratio}
\end{figure}

\subsection{Bound for arbitrary time-varying protocols}

Here, we tackle the case a random system subject to arbitrary time-varying protocols. In this case some meaningful observables, such as work and heat, are not time-antisymmetric in the naive sense of time-reversal as reading the list of states in reverse order, because work and heat depend on the parameters of the time-varying protocol. 

For this reason, we consider the auxiliary configuration space $\Omega = \Omega_\text{forw} \times \Omega_\text{back}$, which is the space of all pairs of paths $(\omega,\omega')$, endowed with the direct product measure $p(\omega,\omega')=p_\text{forw}(\omega) p_\text{back}(\omega')$. Here $p_\text{forw}(\omega)$ evaluates the probability of $\omega$ in the forward protocol, and $p_\text{back}(\omega')$ is the probability computed in the time-reversed protocol.

On $\Omega$ we consider the involution $(\omega,\omega') \mapsto (\overline{\omega}',\overline{\omega})$. In other words the involution reverses \emph{and swaps} the paths. We consider the legitimate observables on $\Omega$ as those that take the form $F(\omega,\omega')=f(\omega)+f'(\omega')$ and are antisymmetric for the involution, which is equivalent to the identity $f(\omega)=-f'(\overline{\omega}) $.  One checks that protocol-dependent thermodynamic variables, such as heat are indeed anti-symmetric for this involution, where in this case $f$ (resp., $f'$) denotes the heat exchanged along the path as computed from the forward (resp., backward) protocol.

We can now apply the bound \eqref{eq:vdb}, only with the linear form $\mean{f}$ occurring in \eqref{prec} now being the sum $\mean{F}=\mean{f}_\text{forw}+\mean{f'}_\text{backw}$ of means according to the forward and backward protocol (similarly for the variance). Moreover, $\mean{\sigma}$ turns out to be 
$$
\mean{\sigma} =\mean{\ln \frac{p_\text{forw}}{\overline{p}_\text{back}}}_\text{forw}+\mean{\ln \frac{p_\text{back}}{  \overline{p}_\text{forw}  }}_\text{back}.
$$
In this way we retrieve the recent result of \cite{pro19}, which is there derived with a large deviation argument. We refer to that paper for a discussion on the meaning and importance of this bound. 

\subsection{Periodic Markovian processes}

We now formulate generalities on periodic Markovian processes, introducing progressively the assumptions of Markovianity on the path level, then periodicity, and finally the construction of a state space. This will be useful to state and prove the Periodic Uncertainty Relation in the next section.

We decompose the configuration space  $\Omega$  as a Cartesian product  $\Omega_0 \times \Omega_1 \times  \ldots \Omega_{N-1}$, so that a global configuration $\omega$ is seen as the concatenation of $N$ local configurations $\omega_0, \omega_1, \ldots, \omega_{N-1}$. Although the formalism applies in principle to any sort of configurations (for instance spin configurations), for consistency with the main text and application to the Thermodynamic Uncertainty Relations, from now on we refer to $\omega$ and $\omega_i$ and as global and local `paths'.

The global probability measure $p(\omega)$ on $\Omega$ naturally projects into marginal probability measures $p_i(\omega_i)$ on each $\Omega_i$, and into marginal pairwise probability measures $p_{ji}(\omega_j,\omega_i)$ on each pair $\Omega_j \times \Omega_i$. The so-called \emph{time-summed observables} on $\Omega$ are those of the form $f=\sum_{i=0}^{N-1} f_i $, where $f_i(\omega_i)$ is an observable on $\Omega_{i}$. We take the legitimate observables on $\Omega$ as those time-summed observables $f$ such that each $f_i$ belongs to the space $\mathcal{F}_i$ of legitimate observables on $\Omega_i$.  The mean of a time-summed observable is the sum of local means $\mean{f} = \sum_{i}  \mean{f_i}_i$.  The mean product of two such observables $g=\sum_j g_j$ and $f=\sum_i f_i$ can be written in terms of scalar products on each $\Omega_j$,
\begin{align} \label{eq:Pij}
\langle g | f \rangle = \sum_{i,j} \sum_{\omega_i,\omega_j} g_j  p_{ji} f_i  = \sum_{i,j} \langle g_j  | P_{i|j} f_i \rangle_j.
\end{align}
Here, we decomposed $p_{ji}$ as $p_j p_{i|j}$, with $p_{i|j}(\omega_j,.)$ the conditional probability measure on $\Omega_i$ given $\omega_j$, and wrote $\sum_{\omega_i}  p_{i|j} f_i = P_{i|j} f_i$ to emphasize that it maps  an observable on $\Omega_i$ to an observable on $\Omega_j$ through a linear conditional mean operator $P_{i|j}$. Note that even if $f_i$ is legitimate, i.e. belongs to $\mathcal{F}_i$, the conditional mean observable $P_{i|j} f_i$ may be an arbitrary square-integrable observable on $\Omega_j$, not necessarily legitimate. Observe that $P_{j|i}=P^*_{i|j}$, where $^*$ denotes the adjunction of linear operators between the Hilbert spaces $\mathcal{L}^2(\Omega_i)$ and $\mathcal{L}^2(\Omega_j)$ equipped with their respective scalar products.

We now introduce the assumption that the sequence $\omega_0,\ldots,\omega_{N-1}$ is a Markov chain.  This implies that $P_{i|j}=P_{i|k} P_{k|j}$ for any $i < k < j$,  and also for any $j < k < i$, which is known as  Chapman-Kolmogorov's equation. Moreover, assume that the dynamics is periodic, which means that all $\Omega_i$ can be taken identical with identical marginals $p_i=p_j$, the joint measures $p_{ji}$ only depend on the difference $i-j$, and the spaces of legitimate observables are identical as well, $\mathcal{F}_i=\mathcal{F}_j$. Then, it is enough to consider $P:=P_{j+1|j}$, from which we compute any $P_{i|j}$ as $P^{i-j}$ if $i \geq j$, and    $(P^*)^{j-i}$ if $i \leq j$, where $P^*$ is the adjoint of $P$ for the scalar product $\langle . | . \rangle_j$ over $\Omega_j$. Note that $P$ is but the usual transition matrix appearing in the master equation associated to the discrete-step Markov chain  $\omega_0,\ldots,\omega_{N-1}$, as we may write the propagation of transient probability measures as 
$$
p(\omega_{j+1})=\sum_{\omega_{j}} p(\omega_{j}) P(\omega_{j}, \omega_{j+1}).
$$
Nevertheless, as we assume periodicity, i.e. stationarity of this Markov chain, the master equation is of little use here, except to notice that $p(\omega_{j})$ must be the dominant left-eigenvector of $P$, of eigenvalue 1. The viewpoint explicited above, and used in \eqref{eq:Pij}, sees $P$ as describing the propagation of the conditional mean of an observable instead of the transient probability measures: this is the `Heisenberg viewpoint' dual to the master equation.

From the knowledge of $P$ we can compute the mean product of any two time-summed observables $g=g_0 + \ldots + g_{N-1}$ and $f=f_0+\ldots+f_{N-1}$ over an arbitrary number $N$ of intervals, as given by \eqref{eq:Pij}, which now becomes
\begin{equation}
\langle g | f \rangle = \sum_i \langle g_i  | f_i \rangle_i + \sum_{i,j:i>j}  \langle g_j  |  P^{i-j} f_i \rangle_j  + \sum_{i,j:i<j}  \langle g_j  |  (P^*)^{j-i} f_i \rangle_j
\end{equation}

or, equivalently:
\begin{equation}\label{eq:gf}
\langle g | f \rangle = \sum_i \langle g_i  | f_i \rangle_i + \sum_{i,j:i>j}  \langle  (P^*)^{i-j} g_j  |  f_i \rangle_i  + \sum_{i,j:i<j}  \langle   P^{j-i} g_j  |  f_i \rangle_i.
\end{equation}

To proceed we exploit Markovianity and periodicity further, as they imply the possibility to define a concept of `state space'. This state space is such that to a path $\omega_i$ we can associate a source state $s(\omega_i)$ and a target state $t(\omega_i)$, with the properties that $s(\omega_{i+1})=t(\omega_i)$ and that $\omega_i$, $\omega_{i+1}$ are independent given the state $t(\omega_i)$. With the knowledge of the probability measure on the paths, one can always build in principle (albeit in a non-unique way) a notion of state complying with these properties, as being a sufficient statistics of the past for the future and conversely \cite{cru89, sha01}. In most applications, the reverse situation occurs, where a natural notion of state is given, from which a notion of path is built as a (discrete or continuous) list of successive states.   
Once a state space $X$ is fixed, together with the source map $s$ and target map $t$, one may
endow a probability measure on $X$ as $p(x)= \sum_{t(\omega_i)=x} p(\omega_i)=  \sum_{s(\omega_{i+1})=x} p(\omega_{i+1})$.
From this we define a Hilbert space of square-summable real observables on the state space $X$. 

We now have natural linear mappings between the path observables and the state observables. 
In particular given an observable $f_i(\omega_i)$, we denote $Sf_i$ the mean of $f_i$ knowing the source state. In other words, $$(Sf_i)(x)= \sum_{\omega_i:s(\omega_i)=x} \frac{p(\omega_i)}{p(x)} f_i(\omega_i).$$ 
In other terms, we can write the operator $S$ as a matrix whose entry  $S(x,\omega_i)$ is $p(\omega_i)/p(x)$ if $x=s(\omega_i)$ and $0$ otherwise. 
The adjoint operator $S^*$ is simply the lifting of a state observable to a path observable: if $h(x)$ is a state observable, then $(S^*h)(\omega_i)=h(s(\omega_i))$. If we think of $S^*$ as a matrix, its entry $S^*(\omega_i, x)$ is $1$ if $s(\omega_i)=x$ and 0 otherwise. This is observed by writing down the identity defining $S^*$, namely $\langle S^* h | f_i \rangle= \langle  h | S f \rangle_X$ for all state observables $h$ and all path observables $f_i$.

Similar considerations apply for $T$, the target conditional mean operator. A trivial observation is that $S1=T1=1$: the constant unit path-observable is mapped to the constant unit state-observable. Another observation is that $SS^* = TT^* = Id_X$, the identity on $\mathcal{L}^2(X)$. Moreover, we have $P=T^*S$ and $P^*= S^* T$. 

With these tools at hand, we can state and prove the Periodic Uncertainty Relation.

\subsection{The Periodic Uncertainty Relation}

We state and prove an abstract version of the Periodic Uncertainty Relation, more general than both \eqref{eq:periodicallN} and the activity bound. Roughly speaking, it states that oftentimes the precision reachable over $N$ periods (for any $N >1$) is less than $N$ times the precision reachable over a single period.
 
We work under the same assumptions as in the previous section. Namely, a global path $\omega$ is a list of local paths $\omega_0, \ldots, \omega_{N-1}$, for which we assume a Markovian and periodic dynamics, and we assume we have chosen a state space $X$. Now suppose that for a certain space of legitimate observables on $\Omega_i$, we find a zero-mean averaging observable $M_i$, i.e. checking $\langle M_i | f_i\rangle=\mean{f_i}$ for every legitimate observable $f_i$.

We now introduce the crucial assumption that $M_i$ is such that $SM_i=TM_i$. Then it is easily checked with the identities derived at the end of the previous section that $PP^*M_i=PM_i$ and $P^*PM_i=P^*M_i$.

Let us get back to the computation of $\langle g | f \rangle$, for $g=g_0 + \ldots + g_{N-1}$ and $f=f_0+\ldots+f_{N-1}$. Recall from \eqref{eq:gf} that the total contribution of $f_i$ in this sum is 
\begin{align} \label{eq:contribgi}
\langle  &P^i g_0 + P^{i-1} g_1 + \ldots + P g_{i-1} + g_i + (P^*) g_{i+1} + \dots \nonumber \\
&+ (P^*)^{N-i-2}g_{N-2} + (P^*)^{N-i-1}g_{N-1} | f_i  \rangle_i.
\end{align}

Assume that we take $g_0=M_i$ and $g_1= \ldots = g_{N-1}=(Id-P)M_i$. Then the contribution of $f_i$ to  $\langle  g | f  \rangle$ reduces to $\langle  M_i | f_i  \rangle_i$ (using among others the fact that $P^*(Id-P)M_i=0$). Thus $g$ is a zero-mean averaging observable over $\Omega$, as $\langle g | f \rangle = \sum_i \langle M_i | f_i \rangle_i = \langle f \rangle$ for every legitimate $f$. 
Therefore the precision over the $N$ periods is bounded by $\langle g |g \rangle$, which develops as

\begin{eqnarray}
\mathfrak{p}_\text{max} &\leq& \langle M_i | M_i \rangle_i + (N-1)\langle (Id-P)M_i | (Id-P)M_i \rangle_i  \nonumber \\
&=&   N \langle M_i | M_i \rangle_i - (N-1) \langle PM_i | PM_i \rangle_i\\
&\leq&  N \langle M_i | M_i \rangle_i
\end{eqnarray}
using $\langle PM_i | PM_i \rangle_i=\langle M_i | P^*PM_i \rangle_i= \langle M_i | P^*M_i \rangle_i=\langle PM_i | M_i \rangle_i= \langle M_i | PM_i \rangle_i$. 
Therefore, $\langle M_i |M_i \rangle_i$ is not only a bound on the precision over one period but also a bound over the precision over $N$ periods, if scaled with a factor $N$.
In particular, if the legitimate averaging observable $m_i$ satisfies 
\begin{equation}\label{eq:SmTm}
Sm_i = Tm_i,
\end{equation}
then the observable $M_i$ given by \eqref{M} also satisfies $SM_i = TM_i$. Moreover $\langle M_i | M_i \rangle_i= \mathfrak{p}_\text{max}(\Omega_i)$.

We now recapitulate our assumptions and formulate the main result. Assume that over a single period of a periodic Markovian system the legitimate averaging observable $m_i$ satisfies \eqref{eq:SmTm}.  Then the precision that can be achieved over $N$ periods, divided by $N$, is less than the precision that can be achieved in a single period: 
\begin{equation}\label{eq:periodicallNSI}
\frac{\mathfrak{p}_\text{max}(\Omega)}{N}   \leq \mathfrak{p}_\text{max}(\Omega_i).
\end{equation} This is our most general statement of the Periodic Uncertainty Relation.

\subsection{The Periodic Uncertainty Relation for time-antisymmetric observables on overdamped Markov processes}

In this section we consider again a periodic Markovian system, and assume that the legitimate observables over one period $\Omega_i$ are the time-antisymmetric observables, for some definition time-reversal, i.e. any involutive symmetry of $\Omega_i$. We also assume that the path on a period (which up to now has been defined as an element of some arbitrary abstract space $\Omega_i$, from which we can derive a source state and a target state) is a discrete or continuous sequence of states of a Markov process, and the time-reversal simply consists in taking this sequence in reverse order. This is the case when the Markov process  models an overdamped system.

To prove \eqref{eq:SmTm} for the legitimate observable $m_i$ defined by \eqref{general-m}, it is enough to show that $S(1-m_i)=T(1-m_i)$, since $S1=T1$. But $1-m_i= \frac{\overline{p}}{p+\overline{p}}$, thus $S(1-m_i)$ evaluated at state $x$ reads
$$
S(1-m_i)=\frac{1}{p(x)}\sum_{\omega_i:s(\omega_i)=x} \frac{p(\omega_i)\overline{p}(\omega_i)}{p(\omega_i)+\overline{p}(\omega_i)}. 
$$ 
As this expression is symmetric for time-reversal, this is also $S(1-m_i)$ evaluated at state $x$.  Therefore we can apply the Periodic Uncertainty Relation. 

\subsection{The activity bound as a Periodic Uncertainty Relation}

We now prove the activity bound \cite{dit18}: the precision of an observable that is the weighted sum of the transitions undergone by a finite-state continuous time Markov chain during an arbitrary time interval is bounded by the mean number of transitions. 

In the first step, we identify the correct $M_i$ for a short time interval $\Delta t$, and notice that $\langle M_i | M_i \rangle_i$ is the expected number of transitions within time $\Delta t$. In order to prove this, let us first consider a general setting (absolutely no assumption on $\Omega$). When the legitimate observables are defined as those that take zero value on a given subset $\Omega_{\text{z}}$, we find that $m$ is the function that takes zero value on  $\Omega_{\text{z}}$ and  unit value on $\Omega \setminus \Omega_{\text{z}}$. Therefore the maximum normalised precision is $1-p(\Omega_{\text{z}})$ and the maximum precision is $p(\Omega_{\text{z}})^{-1}-1$. The zero-mean observable $M$ in the span of $1$ and $m$ is here taking value $1-p(\Omega_{\text{z}})^{-1}$ on $\Omega_{\text{z}}$ and 1 on $\Omega \setminus \Omega_{\text{z}}$, for which we can check indeed $\langle M  | M \rangle=p(\Omega_{\text{z}})^{-1}-1$.

Coming back to the case of stationary finite state Markov chains over a short time interval, we take  $\Omega_{\text{z}}$ as the constant paths, i.e. those where the walker waits without jumping to another state. We  neglect the possibility of multiple transitions in such a short time, therefore  $1-p(\Omega_{\text{z}})$, equal to $\langle M_i  | M_i \rangle_i$ to first order,  is the probability of a proper transition, which is also the mean number of proper transitions, and is proportional to $\Delta t$.  Thus, the optimal $m_i$ assigns a unit weight to all transitions.

In a second step, we verify that $Sm_i=Tm_i$ as requested by \eqref{eq:SmTm}. Indeed $Sm_i$ evaluated at state $x$ is simply the probability to leave the state in the infinitesimal interval, and $Tm_i$ evaluated at state $x$ is the probability of arrival to $x$, which is the same from stationarity. From there the Periodic Uncertainty Relation applies, and the precision available over any time interval is no larger than the expected number of transitions: this is the activity bound.

\subsection{Computing the variance of a time-summed observable in a stationary finite-state continuous-time Markov chain}

We indicate here how to evaluate numerically the variance and covariances of observables for a stationary ergodic finite-state Markov chain over asymptotically large time intervals. This is useful to evaluate the variance of the displacement observable in the kinesin model, as we show in the next section.  A continuous-time Markov chain is often represented by a master equation, or Kolmogorov forward equation, computing the evolution of a transient probability towards stationarity: 
\begin{align}
&\dot{p}(y)= \sum_x p(x) L(x,y), 
&&\text{or}
&&&\dot{p}=  p L
\end{align}
in matrix notation, where $p$ is a row vector and $L$ is the Laplacian matrix encoding the rates: $L(x,y)=w_{x \to y}$, the rate at which the Markov chain, in state $y$, transitions to another state $x$. The diagonal entry is picked so that every row sums to zero: $L(x,x)= - \sum_{y \neq x} w_{x \to y}$, to ensure preservation of total probability. 
To make an explicit link with the periodic case exposed above, we take $\Omega_i$ as the space of paths of some arbitrary duration $\Delta t$. It is then useful to write the discrete-time master equation which propagates the state over an  interval $\Delta t$:

$$
p_{t+\Delta t}= p(t) e^{\Delta t L}. 
$$

As we already observed in last section, the master equation is of little use for a stationary Markov chain, and we prefer the observable viewpoint.
Given a state observable $h$, assigning value $h(x)$ on each state $x$, then $e^{\Delta t L}h$ is a state observable assigning to each $x$ the mean value of $h$ at time $t+\Delta t$ knowing that the state at time $t$ is $x$. The mean value at time $t$ given the state at time $t+\Delta t$ is encoded in the state observable $e^{\Delta t L^*}h$. Here $L^*$ denotes the adjoint of $L$ under the natural scalar product on states, which is defined elementwise with $(L^*)(x,y)=\frac{p(y)}{p(x)}L(y,x)$.
In the same way that the path-to-path operator $P$ factorizes as $P=T^*S$, the state-to-state operator factorizes as $e^{\Delta t L}=ST^*$. Therefore for $k>0$, we can write $P^k=T^* e^{(k-1)\Delta t L} S$.  

We now consider $\langle  g | f \rangle$ in the case where all $f_i$ are zero-mean and identical to one another, and all $g_j$ are zero-mean and identical to one another. In this case, \eqref{eq:contribgi} provides the scaling of the  covariance  $\langle  g | f \rangle$ with $N$:
$$
\lim_{N \to \infty} \frac{\langle  g | f \rangle}{N} = \langle (\sum_{k>0} P^k + \sum_{k>0} (P^*)^k + Id) g_i  | f_i \rangle_i
$$ 
We can rewrite $\sum_{k >0} P^k g_i=T^* (Id - e^{\Delta t L})^{-1} Sg_i$ which in the limit of short time intervals gives $\sum_{k >0} P^kg_i=-T^* (\Delta t L)^{-1} Sg_i$. Note that the inversion of the non-invertible matrix $L$ is not problematic for an ergodic Markov chain, because $L$ is then invertible on the subspace of zero-mean state observables, such as $Sg_i$.  In the limit of short times, it is convenient to consider a time horizon $t_\text{f}$, with $N=t_\text{f}/\Delta t$, and then take $t_\text{f} \to \infty$, as we are interested in asymptotically large times. 
\begin{align}
&\lim_{t_\text{f} \to \infty} \frac{\langle  g | f \rangle}{t_\text{f}} =  \Delta t^{-1} \langle (-T^* (\Delta t L)^{-1} S - S^* (\Delta t L^*)^{-1} T + Id) g_i  | f_i \rangle_i \nonumber
\end{align}
This expression can be processed further by expressing $ \langle T^* (\Delta t L)^{-1} S  g_i  | f_i \rangle_i$ as the state-space scalar product $\Delta t^{-1} \langle L^{-1} S  g_i  | T f_i \rangle_X$, and similarly for  $ \langle S^* (\Delta t L^*)^{-1} T  g_i  | f_i \rangle_i$. Overall, the covariance for asymptotically large times is

\begin{align}\label{eq:covstat}
\lim_{t_\text{f} \to \infty}  \frac{\langle  g | f \rangle}{t_\text{f}} = & -\Delta t^{-2} \langle L^{-1} S  g_i  | T f_i \rangle_X  - \Delta t^{-2} \langle L^{-1} T  g_i  | S f_i \rangle_X + \Delta t^{-1} \langle g_i  | f_i \rangle_i.
\end{align}
which is the main result of this section. 

If we want to evaluate the covariance of non-zero-mean observables, then we may first center the observable by removing the mean. 


\subsection{The kinesin model}

The kinesin model of the main text consists of 2 states connected by 4 reversible transitions.
Each of these four transitions $j=1,\dots, 4$ out of state $x=1,2$ has an associated rate $w_{x e}$ given by \cite{lau07}:
\begin{align}
  w_{11} & = \omega e^{-\epsilon + \theta_a^+ f}  \,, \qquad&
  w_{22} & = \omega e^{-\theta_b^- f}   \,, \nonumber\\
  w_{12} & = \omega' e^{-\epsilon - \theta_a^- f} \,, \qquad&
  w_{22} & = \omega' e^{\theta_b^+ f} \,, \nonumber\\
  w_{13} & = \alpha e^{-\epsilon + \theta_a^+ f} k_0[\textrm{ATP}] \,, \qquad&
  w_{23} & = \alpha e^{-\theta_b^- f} \,, \nonumber\\
  w_{14} & = \alpha' e^{-\epsilon - \theta_a^- f} k_0[\textrm{ATP}]\,, \qquad&
  w_{24} & = \alpha' e^{\theta_b^+ f} \,. \nonumber
\end{align}

The control parameters are the ATP concentration, [ATP], and the external loading force $F<0$, with $f=F \ell/( k_{\text{B}} T)$ being its associated dimensionless work along the length $\ell$. Typical ranges for \emph{in vitro} experiments are $1 \, \mu\text{M} \lesssim \text{[ATP]}  \lesssim 10^3 \, \mu\text{M}$ and $1 \, \text{pN} \lesssim |F|  \lesssim 10 \, \text{pN}$. Other parameters are chosen as in \cite{lau07, dit18}, i.e. extracted from fits of experimental velocity data.

We are interested in computing the variance of the displacement for arbitrarily large times. For that purpose we use \eqref{eq:covstat} for $f_i=g_i$ encoding the centered displacement on each  path over an infinitesimal time $\Delta t$.
The displacement is  $\pm 1$ on each proper transition, and zero on each constant path ($\Delta t$ is short enough to ignore multiple transitions). The mean displacement $\mean{f_i}$ is proportional to $\Delta t$, therefore the centered displacement on each transition can still be taken to $\pm \ell$ up to a negligible transition, and is set to $-\mean{f_i}$ on both constant paths. 
The mean displacement over $\Delta t$ conditioned on the initial state $x=1,2$ is
\begin{align}
Sf_i = \begin{pmatrix}  \Delta t \,c_	1 -\mean{f_i} \\ -  \Delta t \,c_2  -\mean{f_i} \end{pmatrix},
\end{align}
where $\mean{f_i}=  \Delta t [c_1 p(1)- c_2 p(2)]$, $c_x= \ell \sum_{e=1}^4  (-1)^{e-1} k_{xe}  $. The stationary probabilities read $p(1)= a_2/(a_1+a_2)$ and $p(2)=1-p(1)= a_1/(a_1+a_2)$. The mean displacement  over  $\Delta t$ conditioned on the final state $x=1,2$ is 
\begin{align}
Tf_i = \begin{pmatrix}  -  \Delta t  c_2 p(2)/ p(1) -\mean{f_i} \\    \Delta tc_1 p(1)/ p(2) -\mean{f_i} \end{pmatrix}
\end{align}
while the Laplacian  is 
\begin{align}
L= \begin{pmatrix} -a_1 & a_1 \\ a_2 & -a_2 \end{pmatrix},
\end{align}
where  $a_x= \sum_{e=1}^4 w_{xe}$. Therefore the two state-space products are
\begin{align}\label{corr}
-2 \Delta t^{-2} \langle L^{-1} S  f_i  | T f_i \rangle_X =- \frac{[a_1 (c_1 -\mean{f_i})+a_2 (c_2+ \mean{f_i})] [p(1) (c_1+\mean{f_i})+p(2) (c_2-\mean{f_i})]}{ a_1^2+a_2^2}
\end{align}
while the path-space scalar product is
\begin{align}\label{inst}
\Delta t^{-1}\langle f_i  | f_i \rangle_i = \ell^2  (p(1) a_1+ p(2) a_2) =\ell^2  \frac{a_1 a_2}{a_1+a_2}.
\end{align}
Summing together \eqref{corr} and \eqref{inst} we obtain the variance of the displacement.


We checked these results by relying on the large deviation approach \cite{gar09}. The scaled cumulant generating function of any observable $f$ is found by `tilting' the stochastic matrix $L$ by $v_{xe}(q)= e^{q f_{xe}}$ into
\begin{align}
L_f(q) =
\left( \begin{array}{cc}
-\sum_j k_{1e} & \sum_e k_{1e}  v_{1e}(q) \\
\sum_j k_{2e}  v_{2e}(q) & -\sum_j k_{2e}
\end{array} \right),\label{Wq}
\end{align}
and looking for its leading eigenvalue $g(q)$, i.e. the one satisfying $g(0)=0$ . In \eqref{Wq}, $f_{xe}$ defines the observable $f$. For example, the dynamical activity is obtained for $f_{xe}=1\, \forall x,e$, while the motor displacement for $f_{1e}=-f_{2e}$, and $f_{1e}=\Delta$ $(-\Delta)$ with $e$ odd (even). 
The (scaled) mean and the variance of $f$ are calculated as $\langle f \rangle = \partial_q g |_{q=0}$ and $\text{Var}f / t_\text{f}= \partial^2_q g|_{q=0}$, respectively.

\subsection{An example of involution: spin reversal in an Ising system}
Imagine a classical Ising system of $n$ spins $s_i$, in an external magnetic field $h$, equilibrated at inverse temperature $\beta$. The Gibbs probability measure is thus
\begin{align}
p(\{s_i\})\propto e^{- \beta[H_0(\{s_i\}) - h \mathcal{M} ]},
\end{align}
where $H_0(\{s_i\})=H_0(\{-s_i\})$ is the interaction Hamiltonian and  $\mathcal{M}= \sum_{i=1}^n s_i/n$ is the system magnetization.
 Considering spin reversal $\overline{s}_i=-s_i$, entailing the legitimate observables $f(\{s_i\})=-f(\{-s_i\})$,  \eqref{eq:vdb} is an upper bound on the precision of, e.g., the magnetization, taking the form (cf. with \cite{gui16})
\begin{align}
\frac{\langle \mathcal{M} \rangle^2}{\text{Var}\mathcal{M}} \leqslant \frac{e^{\beta h \langle \mathcal{M} \rangle}-1}{2}.
\end{align}

 Beyond the classical case, it is also evident that the same Hilbert uncertainty principle apply to the quantum case. The spin system is then characterized by a density matrix $\rho$ rather than a probability measure, and the mean and second moment of an observable $f$ (now a Hermitian matrix) are computed as the trace of $\rho f$ and $\rho f f^*$ respectively. Now for any subspace of  legitimate observables, one may again find an appropriate maximum precision.   

\end{widetext}

\end{document}